\documentclass[prl,reprint,showpacs,superscriptaddress,longbibliography,amsmath,amssymb,aps] {revtex4-1}
\usepackage{graphicx}
\usepackage{dcolumn}
\usepackage{bm}

\usepackage{afterpage}
\usepackage{float}
\usepackage{placeins}
\usepackage{dsfont}
\usepackage[utf8]{inputenc}
\usepackage{bm}
\usepackage{amssymb}
\usepackage{amsmath}
\usepackage{amsfonts}
\usepackage{graphicx}
\usepackage[usenames,dvipsnames]{xcolor} 
\usepackage{dcolumn}
\usepackage{bbold}

\usepackage[bookmarks=true,colorlinks,citecolor=blue,urlcolor=blue]{hyperref}

\usepackage{bbm}
\usepackage{float}

\usepackage{dcolumn}   
\usepackage{bm}        
\usepackage{filecontents}
\usepackage{lineno}

\usepackage{mathtools}
\usepackage[usenames,dvipsnames]{xcolor} 

\usepackage[export]{adjustbox}
\usepackage{adjustbox}

\usepackage{braket}
\usepackage{tikz}

\newcommand\bs[1]{\ensuremath{\boldsymbol{#1}}}

\newcommand{\eq}[1]{Eq.\thinspace(\ref{#1})}
\newcommand{\fig}[1]{Fig.\thinspace{}\ref{#1}}
\newcommand{\fc}[1]{({#1})}
\newcommand{\figc}[2]{Fig.\thinspace{}\ref{#1}\thinspace{}\fc{#2}}

\begin{document}

\title{Local Probes for Charge-Neutral Edge States in Two-Dimensional Quantum Magnets}
\author{Johannes Feldmeier}
\affiliation{Department of Physics and Institute for Advanced Study, Technical University of Munich, 85748 Garching, Germany}
\affiliation{Munich Center for Quantum Science and Technology (MCQST), Schellingstr. 4, D-80799 M{\"u}nchen, Germany}
\author{Willian Natori}
\affiliation{Blackett Laboratory, Imperial College London, London SW7 2AZ, United Kingdom}
\author{Michael Knap}
\affiliation{Department of Physics and Institute for Advanced Study, Technical University of Munich, 85748 Garching, Germany}
\affiliation{Munich Center for Quantum Science and Technology (MCQST), Schellingstr. 4, D-80799 M{\"u}nchen, Germany}
\author{Johannes Knolle}
\affiliation{Department of Physics and Institute for Advanced Study, Technical University of Munich, 85748 Garching, Germany}
\affiliation{Munich Center for Quantum Science and Technology (MCQST), Schellingstr. 4, D-80799 M{\"u}nchen, Germany}
\affiliation{Blackett Laboratory, Imperial College London, London SW7 2AZ, United Kingdom}
\date{\today}

\begin{abstract}
The bulk-boundary correspondence is a defining feature of topological states of matter. However, for quantum magnets such as spin liquids or topological magnon insulators a direct observation of topological surface states has proven challenging because of the charge-neutral character of the excitations. Here we propose spin-polarized scanning tunneling microscopy as a spin-sensitive local probe to provide direct information about charge neutral topological edge states. We show how their signatures, imprinted in the local structure factor, can be extracted by specifically employing the strengths of existing technologies. As our main example, we determine the dynamical spin correlations of the Kitaev honeycomb model with open boundaries. We show that by contrasting conductance measurements of bulk and edge locations, one can extract direct signatures of the existence of fractionalized excitations and non-trivial topology. The broad applicability of this approach is corroborated by a second example of a kagome topological magnon insulator.
\end{abstract}

\maketitle

\textbf{\textit{Introduction.}}--
The search for topological properties of insulating quantum magnets is an exciting, yet challenging task~\cite{broholm2020quantum,knolle2019field}. While related electronic systems saw a swift verification of the bulk-boundary correspondence~\cite{Hsieh_2008_qsh,Mourik2012_scsc,RevModPhys.83.1057,RevModPhys.82.3045} because surface sensitive probes like angle resolved photoemission spectroscopy (ARPES) and scanning tunneling microscopy (STM) were readily available, similar smoking gun signatures remain elusive for magnetic systems due to the charge-neutral character of spin excitations. One route to address this obstacle leads to spin-sensitive local probes, which have recently been proposed as novel tools for identifying fascinating phases of matter such as quantum spin liquids (QSLs)~\cite{PhysRevB.92.165113,Nieva_2018_noise,chatterjee2019_noise,Balents_2010,aftergood2019probing}.

Moreover, recent technological advances in the fabrication of van-der-Waals heterostructures draw particular attention to magnetic quantum systems in \textit{two} dimensions~\cite{burch2018magnetism,gibertini2019magnetic}. In this context, transport measurements of graphene on top of atomically thin insulating magnets have been employed to measure thermodynamic properties of the magnetic layer~\cite{kim2019micromagnetometry}. Here we propose similar heterostructures for tunneling-based surface-spectroscopy in order to probe magnetic excitations~\cite{Klein2018_probing}. A contender to overcome the abovementioned challenges could thus be provided by spin-polarized scanning tunneling microscopy (SP-STM), which is sensitive to local spin excitations through inelastic tunneling processes~\cite{Pietzsch2001_hysteresis,Bode2007_chiral,Rossier2009_stm,Balatsky2010_stm}. This technique has been employed to characterize arrangements of interacting magnetic atoms, including the resolution of spin wave spectra~\cite{Balashov2006_magnon,Spinelli2014_sw}, and might provide access to localized boundary modes~\cite{Delgado2013_fract}. The most direct application of our proposal may thus be the resolution of edge modes in topological magnon insulators (TMIs), indirect signatures of which have been observed in 2D magnets~\cite{onose2010observation,hirschberger2015thermal,chisnell2015topological,Molina2016_topo,Huang2017_layer,Aguilera2020_magnon}.

\begin{figure}[t]
\begin{center}
\includegraphics[trim={0cm 0cm 0cm 0cm},clip,width=.95\linewidth]{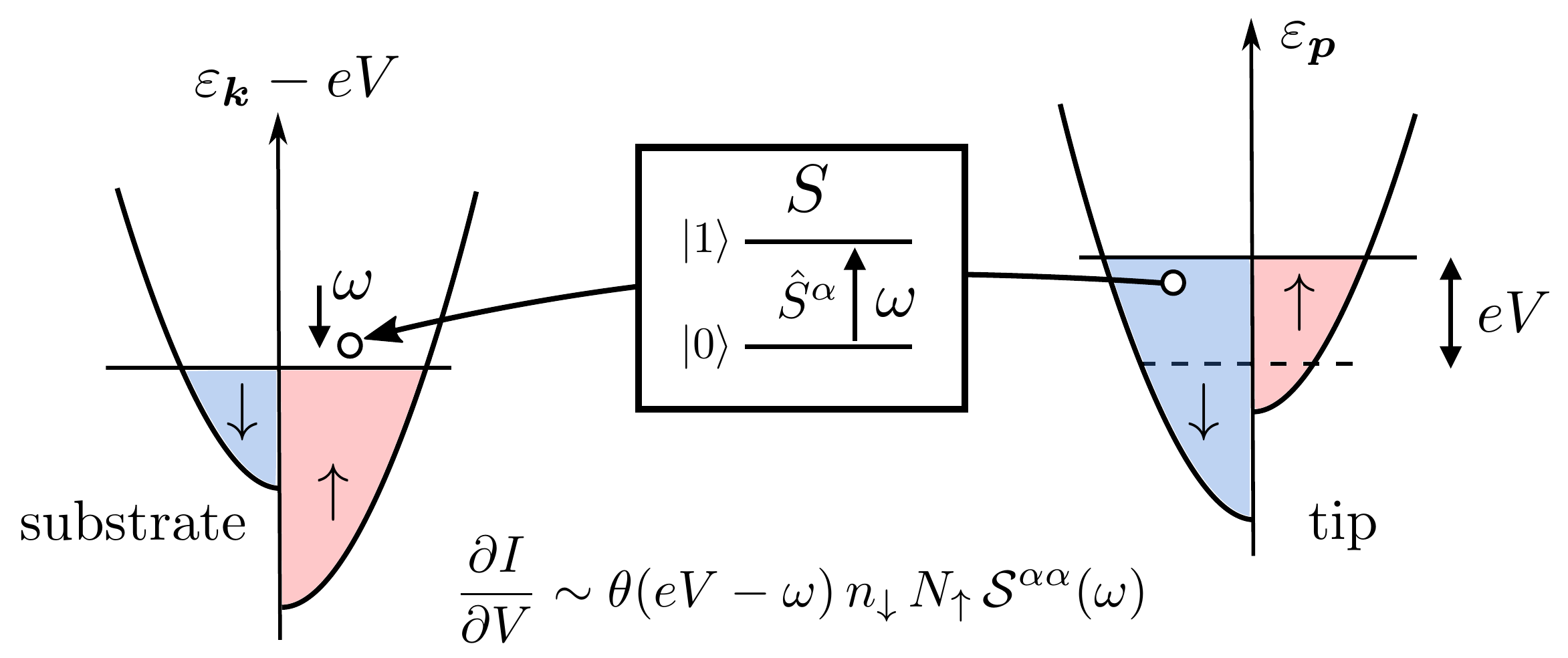}
\caption{\textbf{Spin-polarized scanning tunneling microscopy (SP-STM).} We propose tunneling from a metallic and magnetic substrate to an STM tip via inelastic spin flips of an insulating magnetic layer ($S$) in between. A tunneling electron can excite a mode with energy $\omega$ in $S$ provided the applied bias voltage exceeds this energy. The resulting conductance is proportional to the spin-dependent densities of states in tip, substrate, and the sample $S$, \textit{c.f.} \eq{eq:s1}. Tuning the spin polarization in tip and substrate allows for selectively probing different types of spin excitations in the sample.}
 \label{fig:s1}
\end{center}
\end{figure}

\begin{figure*}[t]
	\centering
	\includegraphics[trim={0cm 0cm 0cm 0cm},clip,width=0.92\linewidth]{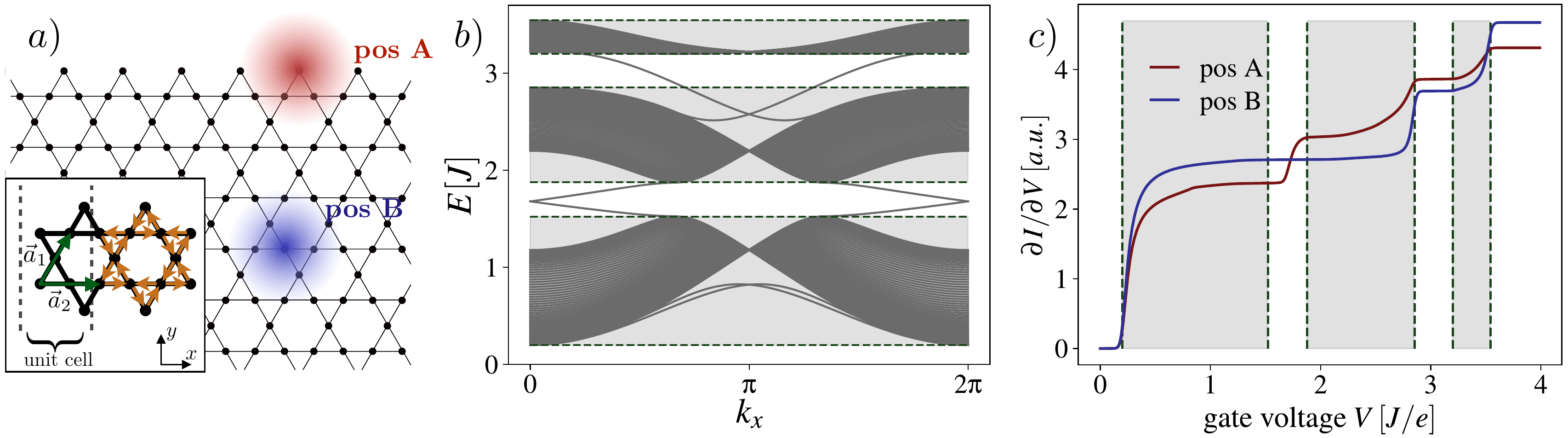}
	\caption{\textbf{Topological Magnon Insulator.} \textbf{a)} The STM tip is placed at the edge (pos A) or over the 2D bulk (pos B) of the Kagome layer, with color gradients indicating the range of the response. For numerical evaluations, a unit cell of $L_y=181$ sites along the $y$-direction is used. Inset: lattice vectors (green) and directions of the DM interaction (orange). \textbf{b)} Energy spectrum for the magnon Hamiltonian with $J=1.0$ and DM-term $D=0.2$, containing edge states within the gaps of the three bulk bands. \textbf{c)} Conductance $\partial I/\partial V$ using \eq{eq:s1}. While the response at tip position B exhibits a flat conductance throughout all band gaps, at the edge (pos A) a finite response within the first gap is acquired, yielding a clear signature for the existence of topological edge magnons.}
	\label{fig:s2}
\end{figure*}

Particular strengths of SP-STM include atomic resolution as well as the ability to investigate anisotropies via selective polarization of tip and substrate, making it in principle well-suited for the study of highly anisotropic Kitaev spin liquids~\cite{kitaev_2006}. Conveniently, one of the prime material candidates~\cite{Jackeli2009_mott,Winter2017_models,Hermanns2018_physics,Takagi2019_concept}, the $\alpha$-$\mathrm{RuCl}_3$ compound, can be exfoliated down to monolayer thickness~\cite{zhou2019possible} and first graphene  heterostructures have been reported~\cite{PhysRevB.100.165426,mashhadi2019spin}. Although this material displays an ordered zig-zag ground state~\cite{Sears2015_magnetic,Johnson2015_monoclinic}, there exists consistent evidence for the onset of a disordered state under the presence of a moderate magnetic field~\cite{Banerjee2016_proximate,Banerjee2017_neutron,Banerjee2018_excitations,Winter2018_probing}. Most strikingly, thermal Hall measurements on bulk samples show a fractional quantization of the thermal conductivity~\cite{Kasahara2018_majorana} indicating the presence of chiral Majorana fermion edge states, a result whose origin is currently under debate~\cite{vinkler2018approximately,ye2018quantization}.

In this work, after a brief summary of SP-STM, we first show that it allows for observing topological magnon edge states of TMIs. As our main result, we then determine qualitative features for potential SP-STM measurements of 2D magnets described by an extended Kitaev honeycomb model. By evaluation of the dynamical spin structure factor on open boundary conditions (OBCs) we find clear signatures associated with the existence of fractionalized gapless edge modes and emergent $\mathbb{Z}_2$ gauge fluxes.

\textbf{\textit{Spin-Polarized STM.}}--
We review some essential aspects of spin-polarized STM, largely based on the works of Refs.~\citep{Rossier2009_stm,Balatsky2010_stm,Bode2003_stm}. The setup is as follows: 
A metallic tip of the STM device (t) is located at a position $\bs{r}=(x,y)$ and at a vertical distance $d$ above a metallic substrate (s). In between, a layer of an insulating spin system (S) is placed on top of the substrate, see \fig{fig:s1}. The Hamiltonian takes the form $\hat{H}=\hat{H}_t+\hat{H}_s+\hat{H}_S+\hat{H}_T$, where $\hat{H}_t=\sum_{\bs{p},\sigma}\varepsilon_{\bs{p},\sigma}\hat{a}^\dagger_{\bs{p},\sigma}\hat{a}_{\bs{p},\sigma}$ and $\hat{H}_s=\sum_{\bs{k},\sigma}\varepsilon_{\bs{k},\sigma}\hat{b}^\dagger_{\bs{k},\sigma}\hat{b}_{\bs{k},\sigma}$ describe the non-interacting electrons in tip and substrate, whose details are not crucial. $\hat{H}_S(\{\hat{\bs{S}}_i\})$ describes the interacting system of spins $\hat{\bs{S}}_i$ at positions $\bs{r}_i$. Finally, $\hat{H}_T$ models the tunneling of electrons between tip and substrate in the presence of an applied bias voltage $V$ via
$
\hat{H}_T = \sum_{\bs{p},\bs{k},\sigma,\sigma^\prime} \Bigl[ \hat{T}^{\sigma\sigma^\prime}_{\bs{r}} \hat{a}^\dagger_{\bs{p},\sigma}\hat{b}^{}_{\bs{k},\sigma^\prime} e^{i\bs{k} \bs{r}+ieVt} + h.c. \Bigr],
$
where $\hat{T}^{\sigma\sigma^\prime}_{\bs{r}}$ depends on the spin system via an exchange coupling, $\hat{T}^{\sigma\sigma^\prime}_{\bs{r}} = t_0\; \delta_{\sigma\sigma^\prime} + \sum_i t_1(\bs{r}-\bs{r}_i)\, \bs{\sigma}_{\sigma\sigma^\prime}\cdot \hat{\bs{S}}_i$. Here, $t_0$ is the bare tunneling rate, while the spin-dependent second term assumes the exponential form $t_1(\bs{r}-\bs{r}_i) = \Gamma_1 e^{-d/d_0}e^{-|\bs{r}-\bs{r}_i|/\lambda}$ with constants $d_0, \lambda$.

Within this setup, we focus on the tunneling conductance $\partial I/\partial V$ due to the spin-dependent contribution. Defining the dynamical structure factor $\mathcal{S}^{\alpha\alpha}_{ij}(t) = \braket{\hat{S}^\alpha_i(t)\hat{S}^\alpha_j(0)}_S = \int d\omega \, e^{-i\omega t} \mathcal{S}_{ij}^{\alpha\alpha}(\omega)$, Fermi's golden rule yields at zero-temperature, see Supp. Mat.~\cite{supplementary},
\begin{equation} \label{eq:s1}
\begin{split}
\frac{\partial I}{\partial V} = \frac{2e^2}{\hbar}\sum_{i,j,\alpha}t_1(\bs{r}-\bs{r}_i)t_1(\bs{r}-\bs{r}_j)
\; c_{\alpha\beta} \, \int_0^{eV} d\omega \, \mathcal{S}_{ij}^{\alpha\beta}(\omega),
\end{split}
\end{equation}
which contains a spin-weight function $c_{\alpha\beta} = \sum_{\sigma,\sigma^\prime} n_\sigma(\varepsilon_F)N_{\sigma^\prime}(\varepsilon_F) \sigma^{\alpha}_{\sigma^\prime \sigma}\sigma^\beta_{\sigma\sigma^\prime}$. Here, the $\sigma^{\alpha}$ are Pauli matrices and $n_\sigma(\varepsilon_F)/N_\sigma(\varepsilon_F)$ are the spin-dependent densities of states at the Fermi level for both tip/substrate. The intuition behind expression \eq{eq:s1} is summarized in \fig{fig:s1}.
Crucially, the prefactors $c_{\alpha\beta}$ depend on the \textit{relative spin-polarization} of tip and substrate. This allows for a controlled selection of spin excitations that are to be probed~\cite{Rossier2009_stm,Balatsky2010_stm}. We highlight three important settings considered in this work:
\textit{(1)} Non-polarized tip and substrate ($n_+=n_-$ and $N_+=N_-$): $c_{\alpha\beta}\sim \delta_{\alpha\beta}$ and independent of $\alpha$.
\textit{(2)} Fully parallel-polarized tip and substrate ($n_-=N_-=0$): $c_{\alpha\beta} \sim \delta_{\alpha,z}\delta_{\beta,z}$, where $z$ was chosen as the common polarization axis.
\textit{(3)} Fully anti-polarized tip and substrate ($n_-=N_+=0$): $c_{\alpha\beta} \sim (1-\delta_{\alpha,z})(1-\delta_{\beta,z})$.

\textbf{\textit{Topological Magnon Insulators.}}--
As a first example, we apply \eq{eq:s1} to topological magnon edge states appearing in TMI-layers. For concreteness, we consider the well known example of a 2D Kagome ferromagnet featuring non-zero Dzyaloshinskii-Moriya (DM) interactions~\cite{Katsura2010_magnonHall,Zhang2013_magnons,malz2019topological}:
\begin{equation} \label{eq:s6}
\hat{H} = \sum_{\braket{nm}} -J\, \bs{S}_n\cdot\bs{S}_m + \bs{D}_{nm}\cdot (\bs{S}_n\times\bs{S}_m)-\bs{h}\cdot\sum_n\bs{S}_n,
\end{equation}
where $\bs{D}_{nm}$ is the DM interaction on the bond $nm$, and $\bs{h}$ is an external magnetic field along $\hat{\bs{z}}||[111]$. Following Ref.~\cite{Zhang2013_magnons}, \eq{eq:s6} can be brought into quadratic spin wave form by applying a standard Holstein-Primakoff approximation, leading to
$
\hat{H}=\sum_{\braket{nm}} b^\dagger_n H_{nm} b^{}_m + \sum_n H_{nn} b^\dagger_nb^{}_n + E_0.
$
Here, $H_{nm}=-S(J+iD)$ along all bonds oriented counter-clockwise within each elementary triangle and $H_{mn}=-S(J-iD)$ accordingly. The diagonal part is given by $H_{nn}=hS+JSM_n$, with $M_n$ the number of nearest neighbors of site $n$, see \figc{fig:s2}{a}. 

On a strip-geometry, $\hat{H}$ can be block-diagonalized with respect to the $k_x$-momentum quantum number such that $\hat{H} = \sum_{k_x} \sum_{l, l^\prime} b^\dagger_l(k_x) \tilde{H}_{l l^\prime}(k_x) b^{}_{l^\prime}(k_x)=\sum_{k_x}\sum_{l}\, \varepsilon_l(k_x) \,\tilde{b}^\dagger_l(k_x)\tilde{b}^{}_l(k_x)$, where $l$ labels the sites along the $y$-direction and the eigenmodes $\tilde{b}^{}_l(k_x)=\sum_{l'} U_{l,l'}(k_x) b^{}_{l'}(k_x)$  are obtained numerically. 
The spectrum $\varepsilon_l(k_x)$ is shown in \figc{fig:s2}{b} and displays edge modes within the bulk gaps between bands with non-zero Chern numbers~\cite{Katsura2010_magnonHall}. The structure factor entering the differential conductance \eq{eq:s1} can be determined simply from its Lehmann representation at finite temperatures. Focusing on the $T=0$ limit, we obtain $\mathcal{S}^{zz}_{l_ml_n}(k_x,\omega) \sim \delta(\omega)$ and $\mathcal{S}^{yy}_{l_ml_n}(k_x,\omega) = \mathcal{S}^{xx}_{l_ml_n}(k_x,\omega)$ with
\begin{equation} \label{eq:s7}
\mathcal{S}^{xx}_{l_ml_n}(k_x,\omega) = \sum_s U_{l_m,s}(k_x) U^*_{l_n,s}(k_x)\, \delta(\omega-\varepsilon_s(k_x)).
\end{equation}
\eq{eq:s7} makes the coupling of the structure factor to the local density of the eigenmodes manifest. Accordingly, $\partial I/ \partial V$, evaluated for an \textit{unpolarized} tip on the boundary of a system containing $181$ sites along the $y$-direction, shows a finite response within the first band gap, see \figc{fig:s2}{c}. We chose $\lambda = 1.0$ (units lattice spacing), which sets the length scale of the tunneling matrix element, and notice that sizable contributions to $\partial I/ \partial V$ arise only from momenta $k_x\lesssim 1/\lambda$~\cite{supplementary}, yielding a finite gap-response from topological magnon edge modes only within the first band gap.

\begin{figure*}[t]
\centering
\includegraphics[trim={0cm 0cm 0cm 0cm},clip,width=0.99\linewidth]{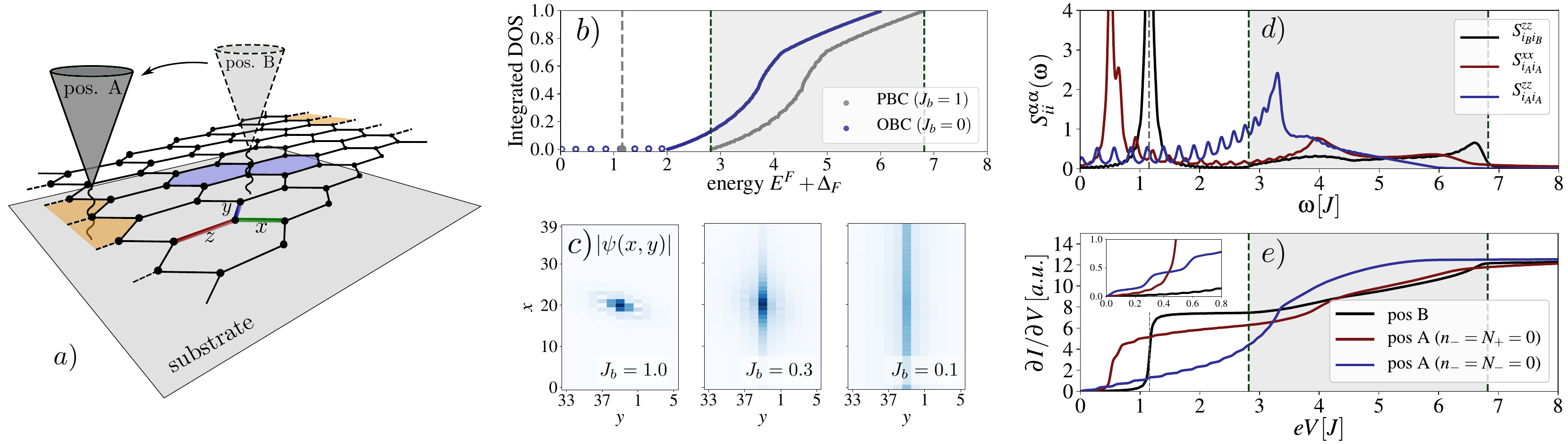}
\caption{\textbf{Kitaev Spin Liquid.} \textbf{a)} Geometry of the STM setup and sketch of the Kitaev model \eq{eq:s2}. OBCs are introduced by setting the strength of the dashed bonds to $J_b\rightarrow 0$. Probing the local spin noise in the bulk requires the creation of a gapped flux-pair (blue). Moving to the boundary, these fluxes are gapless (orange), allowing for the detection of gapless Majorana edge modes. \textbf{b)} Integrated DOS for the fermionic spectrum in the presence of a boundary flux-pair for PBCs ($J_b=1$, gray) and OBCs ($J_b=0$, blue). \textbf{c)} The wave function $\psi(x,y)$ of the fermionic bound state on $40\times 40$ unit cells delocalizes upon reducing the coupling $J_b$ across the boundary. \textbf{d)} Components of the dynamical structure factor for $56\times 56$ unit cells. A finite frequency broadening was introduced for the sharp delta-response from the bound state. \textbf{e)} Conductance for the tip at positions A and B, and different polarizations of tip and substrate. In the bulk, the fermion bound state creates a sharp step within the gap. On the boundary, the step is replaced by a continuum due to the dispersive edge modes, starting from zero bias, see Inset.}
\label{fig:s3}
\end{figure*}

\textbf{\textit{Kitaev Spin Liquid.}}--
We proceed to characterize our main example, the extended Kitaev honeycomb model,
\begin{equation} \label{eq:s2}
\hat{H} = \sum_{\braket{ij}_\alpha} J_\alpha\,\hat{\sigma}^\alpha_i\hat{\sigma}^\alpha_j + K\sum_{\braket{ij}_\alpha,\braket{jk}_\gamma} \hat{\sigma}^\alpha_i \hat{\sigma}^\beta_j \hat{\sigma}^\gamma_k,
\end{equation}
where $\braket{i,j}_\alpha$ denotes nearest neighbors, with $\alpha\in \{x,y,z\}$ labelling the three inequivalent bond types, see \figc{fig:s3}{a} for a schematic picture of the setup.
Following Ref.~\cite{kitaev_2006}, the model can be solved by representing the spin operators $\hat{\sigma}^\alpha_i=i\hat{b}^\alpha_i\hat{c}_i$ in terms of four different Majorana species, resulting in
\begin{equation} \label{eq:s3}
\hat{H} = i \sum_{\braket{ij}_\alpha} J_\alpha \, \hat{u}_{\braket{ij}_\alpha}\, \hat{c}_i \hat{c}_j + iK\sum_{\braket{ij}_\alpha, \braket{jk}_\gamma} \hat{u}_{\braket{ij}_\alpha}\hat{u}_{\braket{jk}_\gamma}\hat{c}_i \hat{c}_k,
\end{equation}
where $\hat{u}_{\braket{ij}_\alpha}=i\hat{b}^\alpha_i\hat{b}^\alpha_j$ are constants of motion with eigenvalues $u_{\braket{ij}_\alpha}=\pm 1$.  There exists a local $\mathbb{Z}_2$ gauge structure with associated plaquette Wilson loops $\hat{W}_p=\prod_{\braket{ij}\in p}\hat{u}_{\braket{ij}_\alpha}$ labelling the gauge sector of the theory. Within a fixed sector of $u_{\braket{ij}\alpha}$'s, \eq{eq:s3} reduces to a Majorana hopping problem.

A convenient description of the model \eq{eq:s3} is obtained by pairing the Majoranas into complex \textit{matter} fermions $\hat{f}^{}_{\bs{r}} = \frac{1}{2}(\hat{c}_{A\bs{r}} + i\,\hat{c}_{B\bs{r}})$ in each unit cell, and \textit{gauge} fermions $\hat{\chi}^{}_{\braket{ij}_\alpha} = \frac{1}{2}(\hat{b}^\alpha_i + i\, \hat{b}^\alpha_j)$ on the bonds, $i \in A$, $j \in B$. 
The $\hat{u}_{\braket{ij}_\alpha} = 2 \hat{\chi}^\dagger_{\braket{ij}_\alpha} \hat{\chi}^{}_{\braket{ij}_\alpha} - 1$ can then be expressed in terms of the gauge fermions, and the ground state is written as $\ket{0}=\ket{M_0}\otimes\ket{F_0}$, with $\ket{M_0}$ the ground state of the matter fermion problem defined by \eq{eq:s3} within the flux-free gauge sector $\ket{F_0}$, for which $W_p=+1$ for all plaquettes.

To obtain OBCs, we choose a line of `weak bonds' around the torus ($z$-bonds w.l.o.g.) whose strength $J_b\rightarrow 0$ vanishes. This results in a degeneracy throughout the many-body spectrum, as the insertion of flux pairs via $u_{\braket{ij}_b}\rightarrow-1$ adjacent to bonds $\braket{ij}_b$ across the boundary comes without energy cost. A  general ground state for OBCs can then be written as
\begin{equation} \label{eq:s4}
\ket{0} = \ket{M_0}\otimes\ket{F}=\ket{M_0}\otimes\ket{F_0}_{bulk}\otimes\ket{F}_b,
\end{equation}
where $\ket{F_0}_{bulk}$ is the flux-free sector of all bulk plaquettes and $\ket{F}_b$ is a general superposition of $2^{L-1}$ different boundary flux sectors for a boundary of length $L$, see Supp. Mat. for more details~\cite{supplementary}.

In order to determine the conductance through \eq{eq:s1}, we have to compute the dynamical structure factor $\mathcal{S}^{\alpha\beta}_{ij}(t) = \braket{0|\hat{\sigma}^\alpha_i(t)\hat{\sigma}^\beta_j(0)|0}$ from a given ground state of \eq{eq:s4}. Following Refs.~\citep{Knolle2014_spinliquid,knolle_2015_fract,knolle_2016_dynamics}, the problem can be reduced to a Majorana quantum quench in the matter sector,
\begin{equation} \label{eq:s5}
\begin{split}
\mathcal{S}^{\alpha\beta}_{ij}(t) &= \braket{M_0|e^{it\hat{H}}\hat{c}_ie^{-it(\hat{H}+\hat{V}_{\braket{il}_\alpha})}\hat{c}_j|M_0} \times \\
& \times \bra{F}(\hat{\chi}^{}_{\braket{il}_\alpha}+\hat{\chi}^\dagger_{\braket{il}_\alpha})(\hat{\chi}^{}_{\braket{jk}_\beta}+\hat{\chi}^\dagger_{\braket{jk}_\beta})\ket{F}.\\
\end{split}
\end{equation}
Here, we chose both $i,j$ on sublattice $A$, and $\hat{V}_{\braket{il}_\alpha}$ is the modification of the Majorana model due to flux insertion  $u_{\braket{il}_\alpha}\rightarrow -u_{\braket{il}_\alpha}$. For bonds $\braket{jk}_\beta$ adjacent to bulk plaquettes, the gauge sector of \eq{eq:s5} reduces to $\mathcal{S}^{\alpha\beta}_{ij}\sim \delta_{\alpha\beta}\,\delta_{ij}$, i.e. the structure factor is ultra-local in the bulk due to the static nature of the gauge field~\cite{baskaran2007exact}. In contrast, bonds $\braket{jk}_\beta=\braket{jk}_b$ across the boundary can acquire longer-range contributions $\mathcal{S}^{\alpha\beta}_{ij}\nsim \delta_{ij}$ due to the superposition $\ket{F}_b$ of boundary fluxes. Nevertheless, while \eq{eq:s5} thus generally depends on the choice of $\ket{F}_b$, the on-site contributions $\mathcal{S}^{\alpha\alpha}_{ii}(t)$ are independent of the chosen state $\ket{0}$, see Supp. Mat.~\cite{supplementary}. Since these contributions dominate the STM response according to \eq{eq:s1}, any choice of $\ket{0}$ will lead to a qualitatively representative conductance $\partial I/ \partial V$. We choose $\ket{F}_b=\ket{F_0}_b$ as flux-free in the following and numerically evaluate \eq{eq:s5} using a Pfaffian approach~\citep{knolle_2015_fract}. In practice, we introduce a small but finite bond-strength $J_b\ll 1$ across the boundary, which provides additional physical insight on the emergence of a Majorana zero mode for $J_b=0$.

Our main results are summarized in \fig{fig:s3}: In panel (b) we show the integrated density of states (DOS) for the matter fermions for $J=1$, $K=0.2$ in a background containing a flux pair adjacent to a weak bond $\braket{ij}_b$ across the boundary. For $J_b=1$ we recover the result for periodic boundaries (PBCs) with an exponentially localized fermion bound state at the flux pair, with an energy $E=\Delta_F + E^F_1 = 1.156J$ (grey dashed), located in the gap below the onset of a continuum band at $E=\Delta_F + E^F_2=2.819J$. Here, $\Delta_F=0.819$ is the two-flux gap in the bulk and $E^F_{1/2}$ the first/second eigenstate of the matter model. As we decrease $J_b$, \fig{fig:s3}{c} shows how the bound state delocalizes along the boundary, eventually turning into a zero mode. This is reflected in the DOS by an emerging continuum of in-gap states (blue line and circles in panel b), corresponding to a dispersive chiral Majorana edge mode, as well as a vanishing flux gap.

Crucially, these spectral  properties of Majorana-flux bound states and the chiral Majorana edge modes are directly reflected in the local structure factor, displayed in \figc{fig:s3}{d} and evaluated for $J_b=0.01J$: $\mathcal{S}^{\alpha\alpha}_{i_Bi_B}(\omega)$ at site $i_B$ in the bulk (see \figc{fig:s3}{a}) reflects the spectrum of PBCs via a sole, sharp contribution at the bound state energy and a broad continuum at higher frequencies. Note, similar signatures for the Majorana-flux bound state have been very recently predicted for planar tunneling spectroscopy~\cite{carrega2020tunneling}. In contrast, the component  $\mathcal{S}^{zz}_{i_Ai_A}(\omega)$ (blue) at a boundary site $i_A$ contains no sharp contribution and instead exhibits a spectral response throughout the former excitation gap. This demonstrates that the structure factor couples directly to the gapless Majorana edge mode. The component $\mathcal{S}^{xx}_{i_Ai_A}(\omega)$ involves the creation of a single bulk flux and has a sharp onset at a reduced flux gap $\Delta_F=0.499J$, above which dispersive edge modes give a finite in-gap response.

The conductance derived from these results, see \figc{fig:s3}{e}, is evaluated via \eq{eq:s1} for a small $\lambda = 0.1$ (units lattice constant), essentially focusing on the on-site response. For tip position B, the polarizations entering $c_{\alpha\beta}$ do not have qualitative effects due to symmetry of the bulk structure factor. The resulting conductance features a sharp step at the bound state energy. At the boundary (position A), the conductance varies drastically with changing $c_{\alpha\beta}$: An anti-polarized tip captures the features of $\mathcal{S}^{xx}_{i_Ai_A}(\omega)=\mathcal{S}^{yy}_{i_Ai_A}(\omega)$ through a sharp step for a bias voltage matching the reduced flux gap, followed by smaller steps due to edge states. These smaller steps merge into a continuum in the thermodynamic limit. Note, contrasting the response of the bulk and edge modes even enables the measurement of single flux and nearest-neighbor flux-pair energies. The latter has a value less than twice the single flux energy because of Majorana induced interactions.  
Finally, for a parallel-polarized setting, where $c_{\alpha\beta}$ exclusively picks up the $\mathcal{S}^{zz}_{i_Ai_A}(\omega)$-component, the flux excitation has no effect, resulting in an approximately linear increase of $\partial I/\partial V$ throughout the bulk-gap, in particular also at zero bias, providing a clear signature of the chiral Majorana edge modes.

\textbf{\textit{Conclusions \& Outlook.}}--
In this work, we proposed tunable SP-STM measurements for probing site-local and spin-anisotropic characteristics of 2D quantum magnets. In particular, we obtained characteristic tunneling signatures of topological magnon edge modes for TMIs. As our main result, we established that fractionalized vison and Majorana fermion excitations of the Kitaev QSLs can be measured via SP-STM by contrasting bulk and boundary measurements. 
Our analysis further demonstrates the direct coupling of the spin structure factor to the Majorana correlation function on the system boundary, leading to contributions beyond nearest neighbor separation due to a modified flux selection rule.

In the future, it would be desirable to investigate whether such longer range correlations can be probed by spin noise spectroscopy measurements, possibly providing an even more direct probe of the chiral nature of the Majorana edge modes. Furthermore, the gapless nature of the edge response in the Kitaev model could open a route for a larger variety of spin-sensitive spectroscopy tools. In particular, nitrogen-vacancy magnetometry, typically operating on energy scales of up to $\sim 100 \mathrm{GHz}$~\cite{Casola2018_nv}, well below the typical values of exchange parameters of candidate materials in the $\mathrm{THz}$-regime, might be used to further characterize 1D edge physics in several bulk Kitaev materials, i.e. $\alpha$-RuCl$_3$~\cite{Jackeli2009_mott,Winter2017_models,Hermanns2018_physics,Takagi2019_concept}. 
In conclusion, we have established the potential of local SP-STM probes for confirming and qualitatively characterizing TMI and QSL physics. The observation of unambigous signatures of topological magnon edge modes for the former, and magnetic Majorana fermions as well as gauge flux excitations for the latter, would provide a crucial step towards the long time goal of their controlled manipulation.

\textbf{Acknowledgements.}-- We thank C. Kuhlenkamp and A. Schuckert for insightful discussions. J.K. would like to thank A. Banerjee, M. Burghard, J.C.S. Davis, R. Moessner, T. Oka, M. Udagawa, P. Wahl and especially Y. Matsuda for engaging discussions.  We acknowledge support from the Imperial-TUM flagship partnership, the Royal Society via a Newton International Fellowship through project NIF-R1-181696, the Technical University of Munich - Institute for Advanced Study, funded by the German Excellence Initiative, the European Union FP7 under grant agreement 291763, the Deutsche Forschungsgemeinschaft (DFG, German Research Foundation) under Germany's Excellence Strategy--EXC-2111--390814868, the European Research Council (ERC) under the European Union's Horizon 2020 research and innovation programme (grant agreement No. 851161), from DFG grant No. KN1254/1-1, No. KN1254/1-2, and DFG TRR80 (Project F8).

\bibliography{stm_lit}

\begin{thebibliography}{58}%
\makeatletter
\providecommand \@ifxundefined [1]{%
 \@ifx{#1\undefined}
}%
\providecommand \@ifnum [1]{%
 \ifnum #1\expandafter \@firstoftwo
 \else \expandafter \@secondoftwo
 \fi
}%
\providecommand \@ifx [1]{%
 \ifx #1\expandafter \@firstoftwo
 \else \expandafter \@secondoftwo
 \fi
}%
\providecommand \natexlab [1]{#1}%
\providecommand \enquote  [1]{``#1''}%
\providecommand \bibnamefont  [1]{#1}%
\providecommand \bibfnamefont [1]{#1}%
\providecommand \citenamefont [1]{#1}%
\providecommand \href@noop [0]{\@secondoftwo}%
\providecommand \href [0]{\begingroup \@sanitize@url \@href}%
\providecommand \@href[1]{\@@startlink{#1}\@@href}%
\providecommand \@@href[1]{\endgroup#1\@@endlink}%
\providecommand \@sanitize@url [0]{\catcode `\\12\catcode `\$12\catcode
  `\&12\catcode `\#12\catcode `\^12\catcode `\_12\catcode `\%12\relax}%
\providecommand \@@startlink[1]{}%
\providecommand \@@endlink[0]{}%
\providecommand \url  [0]{\begingroup\@sanitize@url \@url }%
\providecommand \@url [1]{\endgroup\@href {#1}{\urlprefix }}%
\providecommand \urlprefix  [0]{URL }%
\providecommand \Eprint [0]{\href }%
\providecommand \doibase [0]{http://dx.doi.org/}%
\providecommand \selectlanguage [0]{\@gobble}%
\providecommand \bibinfo  [0]{\@secondoftwo}%
\providecommand \bibfield  [0]{\@secondoftwo}%
\providecommand \translation [1]{[#1]}%
\providecommand \BibitemOpen [0]{}%
\providecommand \bibitemStop [0]{}%
\providecommand \bibitemNoStop [0]{.\EOS\space}%
\providecommand \EOS [0]{\spacefactor3000\relax}%
\providecommand \BibitemShut  [1]{\csname bibitem#1\endcsname}%
\let\auto@bib@innerbib\@empty
\bibitem [{\citenamefont {Broholm}\ \emph {et~al.}(2020)\citenamefont
  {Broholm}, \citenamefont {Cava}, \citenamefont {Kivelson}, \citenamefont
  {Nocera}, \citenamefont {Norman},\ and\ \citenamefont
  {Senthil}}]{broholm2020quantum}%
  \BibitemOpen
  \bibfield  {author} {\bibinfo {author} {\bibfnamefont {C.}~\bibnamefont
  {Broholm}}, \bibinfo {author} {\bibfnamefont {R.~J.}\ \bibnamefont {Cava}},
  \bibinfo {author} {\bibfnamefont {S.~A.}\ \bibnamefont {Kivelson}}, \bibinfo
  {author} {\bibfnamefont {D.~G.}\ \bibnamefont {Nocera}}, \bibinfo {author}
  {\bibfnamefont {M.~R.}\ \bibnamefont {Norman}}, \ and\ \bibinfo {author}
  {\bibfnamefont {T.}~\bibnamefont {Senthil}},\ }\bibfield  {title} {\enquote
  {\bibinfo {title} {{Quantum spin liquids}},}\ }\href {\doibase
  10.1126/science.aay0668} {\bibfield  {journal} {\bibinfo  {journal}
  {Science}\ }\textbf {\bibinfo {volume} {367}} (\bibinfo {year} {2020}),\
  10.1126/science.aay0668}\BibitemShut {NoStop}%
\bibitem [{\citenamefont {Knolle}\ and\ \citenamefont
  {Moessner}(2019)}]{knolle2019field}%
  \BibitemOpen
  \bibfield  {author} {\bibinfo {author} {\bibfnamefont {J.}~\bibnamefont
  {Knolle}}\ and\ \bibinfo {author} {\bibfnamefont {R.}~\bibnamefont
  {Moessner}},\ }\bibfield  {title} {\enquote {\bibinfo {title} {{A Field Guide
  to Spin Liquids}},}\ }\href {\doibase
  10.1146/annurev-conmatphys-031218-013401} {\bibfield  {journal} {\bibinfo
  {journal} {Annual Review of Condensed Matter Physics}\ }\textbf {\bibinfo
  {volume} {10}},\ \bibinfo {pages} {451--472} (\bibinfo {year}
  {2019})}\BibitemShut {NoStop}%
\bibitem [{\citenamefont {{Hsieh, D. and Qian, D. and Wray, L. and Xia, Y. and
  Hor, Y. S. and Cava, R. J. and Hasan, M. Z.}}(2008)}]{Hsieh_2008_qsh}%
  \BibitemOpen
  \bibfield  {author} {\bibinfo {author} {\bibnamefont {{Hsieh, D. and Qian, D.
  and Wray, L. and Xia, Y. and Hor, Y. S. and Cava, R. J. and Hasan, M. Z.}}},\
  }\bibfield  {title} {\enquote {\bibinfo {title} {{A topological Dirac
  insulator in a quantum spin Hall phase}},}\ }\href {\doibase
  10.1038/nature06843} {\bibfield  {journal} {\bibinfo  {journal} {Nature}\
  }\textbf {\bibinfo {volume} {452}},\ \bibinfo {pages} {970--974} (\bibinfo
  {year} {2008})}\BibitemShut {NoStop}%
\bibitem [{\citenamefont {Mourik}\ \emph {et~al.}(2012)\citenamefont {Mourik},
  \citenamefont {Zuo}, \citenamefont {Frolov}, \citenamefont {Plissard},
  \citenamefont {Bakkers},\ and\ \citenamefont
  {Kouwenhoven}}]{Mourik2012_scsc}%
  \BibitemOpen
  \bibfield  {author} {\bibinfo {author} {\bibfnamefont {V.}~\bibnamefont
  {Mourik}}, \bibinfo {author} {\bibfnamefont {K.}~\bibnamefont {Zuo}},
  \bibinfo {author} {\bibfnamefont {S.~M.}\ \bibnamefont {Frolov}}, \bibinfo
  {author} {\bibfnamefont {S.~R.}\ \bibnamefont {Plissard}}, \bibinfo {author}
  {\bibfnamefont {E.~P. A.~M.}\ \bibnamefont {Bakkers}}, \ and\ \bibinfo
  {author} {\bibfnamefont {L.~P.}\ \bibnamefont {Kouwenhoven}},\ }\bibfield
  {title} {\enquote {\bibinfo {title} {{Signatures of Majorana Fermions in
  Hybrid Superconductor-Semiconductor Nanowire Devices}},}\ }\href {\doibase
  10.1126/science.1222360} {\bibfield  {journal} {\bibinfo  {journal}
  {Science}\ }\textbf {\bibinfo {volume} {336}},\ \bibinfo {pages} {1003--1007}
  (\bibinfo {year} {2012})}\BibitemShut {NoStop}%
\bibitem [{\citenamefont {Qi}\ and\ \citenamefont
  {Zhang}(2011)}]{RevModPhys.83.1057}%
  \BibitemOpen
  \bibfield  {author} {\bibinfo {author} {\bibfnamefont {Xiao-Liang}\
  \bibnamefont {Qi}}\ and\ \bibinfo {author} {\bibfnamefont {Shou-Cheng}\
  \bibnamefont {Zhang}},\ }\bibfield  {title} {\enquote {\bibinfo {title}
  {Topological insulators and superconductors},}\ }\href {\doibase
  10.1103/RevModPhys.83.1057} {\bibfield  {journal} {\bibinfo  {journal} {Rev.
  Mod. Phys.}\ }\textbf {\bibinfo {volume} {83}},\ \bibinfo {pages}
  {1057--1110} (\bibinfo {year} {2011})}\BibitemShut {NoStop}%
\bibitem [{\citenamefont {Hasan}\ and\ \citenamefont
  {Kane}(2010)}]{RevModPhys.82.3045}%
  \BibitemOpen
  \bibfield  {author} {\bibinfo {author} {\bibfnamefont {M.~Z.}\ \bibnamefont
  {Hasan}}\ and\ \bibinfo {author} {\bibfnamefont {C.~L.}\ \bibnamefont
  {Kane}},\ }\bibfield  {title} {\enquote {\bibinfo {title} {Colloquium:
  Topological insulators},}\ }\href {\doibase 10.1103/RevModPhys.82.3045}
  {\bibfield  {journal} {\bibinfo  {journal} {Rev. Mod. Phys.}\ }\textbf
  {\bibinfo {volume} {82}},\ \bibinfo {pages} {3045--3067} (\bibinfo {year}
  {2010})}\BibitemShut {NoStop}%
\bibitem [{\citenamefont {Chatterjee}\ and\ \citenamefont
  {Sachdev}(2015)}]{PhysRevB.92.165113}%
  \BibitemOpen
  \bibfield  {author} {\bibinfo {author} {\bibfnamefont {Shubhayu}\
  \bibnamefont {Chatterjee}}\ and\ \bibinfo {author} {\bibfnamefont {Subir}\
  \bibnamefont {Sachdev}},\ }\bibfield  {title} {\enquote {\bibinfo {title}
  {Probing excitations in insulators via injection of spin currents},}\ }\href
  {\doibase 10.1103/PhysRevB.92.165113} {\bibfield  {journal} {\bibinfo
  {journal} {Phys. Rev. B}\ }\textbf {\bibinfo {volume} {92}},\ \bibinfo
  {pages} {165113} (\bibinfo {year} {2015})}\BibitemShut {NoStop}%
\bibitem [{\citenamefont {Rodriguez-Nieva}\ \emph {et~al.}(2018)\citenamefont
  {Rodriguez-Nieva}, \citenamefont {Agarwal}, \citenamefont {Giamarchi},
  \citenamefont {Halperin}, \citenamefont {Lukin},\ and\ \citenamefont
  {Demler}}]{Nieva_2018_noise}%
  \BibitemOpen
  \bibfield  {author} {\bibinfo {author} {\bibfnamefont {J.~F.}\ \bibnamefont
  {Rodriguez-Nieva}}, \bibinfo {author} {\bibfnamefont {K.}~\bibnamefont
  {Agarwal}}, \bibinfo {author} {\bibfnamefont {T.}~\bibnamefont {Giamarchi}},
  \bibinfo {author} {\bibfnamefont {B.~I.}\ \bibnamefont {Halperin}}, \bibinfo
  {author} {\bibfnamefont {M.~D.}\ \bibnamefont {Lukin}}, \ and\ \bibinfo
  {author} {\bibfnamefont {E.}~\bibnamefont {Demler}},\ }\bibfield  {title}
  {\enquote {\bibinfo {title} {{Probing one-dimensional systems via noise
  magnetometry with single spin qubits}},}\ }\href {\doibase
  10.1103/PhysRevB.98.195433} {\bibfield  {journal} {\bibinfo  {journal} {Phys.
  Rev. B}\ }\textbf {\bibinfo {volume} {98}},\ \bibinfo {pages} {195433}
  (\bibinfo {year} {2018})}\BibitemShut {NoStop}%
\bibitem [{\citenamefont {Chatterjee}\ \emph {et~al.}(2019)\citenamefont
  {Chatterjee}, \citenamefont {Rodriguez-Nieva},\ and\ \citenamefont
  {Demler}}]{chatterjee2019_noise}%
  \BibitemOpen
  \bibfield  {author} {\bibinfo {author} {\bibfnamefont {S.}~\bibnamefont
  {Chatterjee}}, \bibinfo {author} {\bibfnamefont {J.~F.}\ \bibnamefont
  {Rodriguez-Nieva}}, \ and\ \bibinfo {author} {\bibfnamefont {E.}~\bibnamefont
  {Demler}},\ }\bibfield  {title} {\enquote {\bibinfo {title} {{Diagnosing
  phases of magnetic insulators via noise magnetometry with spin qubits}},}\
  }\href {\doibase 10.1103/PhysRevB.99.104425} {\bibfield  {journal} {\bibinfo
  {journal} {Phys. Rev. B}\ }\textbf {\bibinfo {volume} {99}},\ \bibinfo
  {pages} {104425} (\bibinfo {year} {2019})}\BibitemShut {NoStop}%
\bibitem [{\citenamefont {Balents}(2010)}]{Balents_2010}%
  \BibitemOpen
  \bibfield  {author} {\bibinfo {author} {\bibfnamefont {L.}~\bibnamefont
  {Balents}},\ }\bibfield  {title} {\enquote {\bibinfo {title} {{Spin liquids
  in frustrated magnets}},}\ }\href {\doibase 10.1038/nature08917} {\bibfield
  {journal} {\bibinfo  {journal} {Nature}\ }\textbf {\bibinfo {volume} {464}},\
  \bibinfo {pages} {199--208} (\bibinfo {year} {2010})}\BibitemShut {NoStop}%
\bibitem [{\citenamefont {Aftergood}\ and\ \citenamefont
  {Takei}(2019)}]{aftergood2019probing}%
  \BibitemOpen
  \bibfield  {author} {\bibinfo {author} {\bibfnamefont {J.}~\bibnamefont
  {Aftergood}}\ and\ \bibinfo {author} {\bibfnamefont {S.}~\bibnamefont
  {Takei}},\ }\href@noop {} {\enquote {\bibinfo {title} {{Probing quantum spin
  liquids in equilibrium using the inverse spin Hall effect}},}\ } (\bibinfo
  {year} {2019}),\ \Eprint {http://arxiv.org/abs/1910.08610} {arXiv:1910.08610}
  \BibitemShut {NoStop}%
\bibitem [{\citenamefont {Burch}\ \emph {et~al.}(2018)\citenamefont {Burch},
  \citenamefont {Mandrus},\ and\ \citenamefont {Park}}]{burch2018magnetism}%
  \BibitemOpen
  \bibfield  {author} {\bibinfo {author} {\bibfnamefont {K.~S.}\ \bibnamefont
  {Burch}}, \bibinfo {author} {\bibfnamefont {D.}~\bibnamefont {Mandrus}}, \
  and\ \bibinfo {author} {\bibfnamefont {J.-G.}\ \bibnamefont {Park}},\
  }\bibfield  {title} {\enquote {\bibinfo {title} {{Magnetism in
  two-dimensional van der Waals materials}},}\ }\href {\doibase
  10.1038/s41586-018-0631-z} {\bibfield  {journal} {\bibinfo  {journal}
  {Nature}\ }\textbf {\bibinfo {volume} {563}},\ \bibinfo {pages} {47--52}
  (\bibinfo {year} {2018})}\BibitemShut {NoStop}%
\bibitem [{\citenamefont {Gibertini}\ \emph {et~al.}(2019)\citenamefont
  {Gibertini}, \citenamefont {Koperski}, \citenamefont {Morpurgo},\ and\
  \citenamefont {Novoselov}}]{gibertini2019magnetic}%
  \BibitemOpen
  \bibfield  {author} {\bibinfo {author} {\bibfnamefont {M.}~\bibnamefont
  {Gibertini}}, \bibinfo {author} {\bibfnamefont {M.}~\bibnamefont {Koperski}},
  \bibinfo {author} {\bibfnamefont {A.~F.}\ \bibnamefont {Morpurgo}}, \ and\
  \bibinfo {author} {\bibfnamefont {K.~S.}\ \bibnamefont {Novoselov}},\
  }\bibfield  {title} {\enquote {\bibinfo {title} {{Magnetic 2D materials and
  heterostructures}},}\ }\href {\doibase 10.1038/s41565-019-0438-6} {\bibfield
  {journal} {\bibinfo  {journal} {Nature Nanotechnology}\ }\textbf {\bibinfo
  {volume} {14}},\ \bibinfo {pages} {408--419} (\bibinfo {year}
  {2019})}\BibitemShut {NoStop}%
\bibitem [{\citenamefont {Kim}\ \emph {et~al.}(2019)\citenamefont {Kim},
  \citenamefont {Kumaravadivel}, \citenamefont {Birkbeck}, \citenamefont
  {Kuang}, \citenamefont {Xu}, \citenamefont {Hopkinson}, \citenamefont
  {Knolle}, \citenamefont {McClarty}, \citenamefont {Berdyugin}, \citenamefont
  {Ben~Shalom},\ and\ \citenamefont {et~al.}}]{kim2019micromagnetometry}%
  \BibitemOpen
  \bibfield  {author} {\bibinfo {author} {\bibfnamefont {M.}~\bibnamefont
  {Kim}}, \bibinfo {author} {\bibfnamefont {P.}~\bibnamefont {Kumaravadivel}},
  \bibinfo {author} {\bibfnamefont {J.}~\bibnamefont {Birkbeck}}, \bibinfo
  {author} {\bibfnamefont {W.}~\bibnamefont {Kuang}}, \bibinfo {author}
  {\bibfnamefont {S.~G.}\ \bibnamefont {Xu}}, \bibinfo {author} {\bibfnamefont
  {D.~G.}\ \bibnamefont {Hopkinson}}, \bibinfo {author} {\bibfnamefont
  {J.}~\bibnamefont {Knolle}}, \bibinfo {author} {\bibfnamefont {P.~A.}\
  \bibnamefont {McClarty}}, \bibinfo {author} {\bibfnamefont {A.~I.}\
  \bibnamefont {Berdyugin}}, \bibinfo {author} {\bibfnamefont {M.}~\bibnamefont
  {Ben~Shalom}}, \ and\ \bibinfo {author} {\bibnamefont {et~al.}},\ }\bibfield
  {title} {\enquote {\bibinfo {title} {{Micromagnetometry of two-dimensional
  ferromagnets}},}\ }\href {\doibase 10.1038/s41928-019-0302-6} {\bibfield
  {journal} {\bibinfo  {journal} {Nature Electronics}\ }\textbf {\bibinfo
  {volume} {2}},\ \bibinfo {pages} {457--463} (\bibinfo {year}
  {2019})}\BibitemShut {NoStop}%
\bibitem [{\citenamefont {Klein}\ \emph {et~al.}(2018)\citenamefont {Klein},
  \citenamefont {MacNeill}, \citenamefont {Lado}, \citenamefont {Soriano},
  \citenamefont {Navarro-Moratalla}, \citenamefont {Watanabe}, \citenamefont
  {Taniguchi}, \citenamefont {Manni}, \citenamefont {Canfield}, \citenamefont
  {Fern{\'a}ndez-Rossier},\ and\ \citenamefont
  {Jarillo-Herrero}}]{Klein2018_probing}%
  \BibitemOpen
  \bibfield  {author} {\bibinfo {author} {\bibfnamefont {D.~R.}\ \bibnamefont
  {Klein}}, \bibinfo {author} {\bibfnamefont {D.}~\bibnamefont {MacNeill}},
  \bibinfo {author} {\bibfnamefont {J.~L.}\ \bibnamefont {Lado}}, \bibinfo
  {author} {\bibfnamefont {D.}~\bibnamefont {Soriano}}, \bibinfo {author}
  {\bibfnamefont {E.}~\bibnamefont {Navarro-Moratalla}}, \bibinfo {author}
  {\bibfnamefont {K.}~\bibnamefont {Watanabe}}, \bibinfo {author}
  {\bibfnamefont {T.}~\bibnamefont {Taniguchi}}, \bibinfo {author}
  {\bibfnamefont {S.}~\bibnamefont {Manni}}, \bibinfo {author} {\bibfnamefont
  {P.}~\bibnamefont {Canfield}}, \bibinfo {author} {\bibfnamefont
  {J.}~\bibnamefont {Fern{\'a}ndez-Rossier}}, \ and\ \bibinfo {author}
  {\bibfnamefont {P.}~\bibnamefont {Jarillo-Herrero}},\ }\bibfield  {title}
  {\enquote {\bibinfo {title} {{Probing magnetism in 2D van der Waals
  crystalline insulators via electron tunneling}},}\ }\href {\doibase
  10.1126/science.aar3617} {\bibfield  {journal} {\bibinfo  {journal}
  {Science}\ }\textbf {\bibinfo {volume} {360}},\ \bibinfo {pages} {1218--1222}
  (\bibinfo {year} {2018})}\BibitemShut {NoStop}%
\bibitem [{\citenamefont {Pietzsch}\ \emph {et~al.}(2001)\citenamefont
  {Pietzsch}, \citenamefont {Kubetzka}, \citenamefont {Bode},\ and\
  \citenamefont {Wiesendanger}}]{Pietzsch2001_hysteresis}%
  \BibitemOpen
  \bibfield  {author} {\bibinfo {author} {\bibfnamefont {O.}~\bibnamefont
  {Pietzsch}}, \bibinfo {author} {\bibfnamefont {A.}~\bibnamefont {Kubetzka}},
  \bibinfo {author} {\bibfnamefont {M.}~\bibnamefont {Bode}}, \ and\ \bibinfo
  {author} {\bibfnamefont {R.}~\bibnamefont {Wiesendanger}},\ }\bibfield
  {title} {\enquote {\bibinfo {title} {{Observation of Magnetic Hysteresis at
  the Nanometer Scale by Spin-Polarized Scanning Tunneling Spectroscopy}},}\
  }\href {\doibase 10.1126/science.1060513} {\bibfield  {journal} {\bibinfo
  {journal} {Science}\ }\textbf {\bibinfo {volume} {292}},\ \bibinfo {pages}
  {2053--2056} (\bibinfo {year} {2001})}\BibitemShut {NoStop}%
\bibitem [{\citenamefont {Bode}\ \emph {et~al.}(2007)\citenamefont {Bode},
  \citenamefont {Heide}, \citenamefont {von Bergmann}, \citenamefont
  {Ferriani}, \citenamefont {Heinze}, \citenamefont {Bihlmayer}, \citenamefont
  {Kubetzka}, \citenamefont {Pietzsch}, \citenamefont {Bl{\"u}gel},\ and\
  \citenamefont {Wiesendanger}}]{Bode2007_chiral}%
  \BibitemOpen
  \bibfield  {author} {\bibinfo {author} {\bibfnamefont {M.}~\bibnamefont
  {Bode}}, \bibinfo {author} {\bibfnamefont {M.}~\bibnamefont {Heide}},
  \bibinfo {author} {\bibfnamefont {K.}~\bibnamefont {von Bergmann}}, \bibinfo
  {author} {\bibfnamefont {P.}~\bibnamefont {Ferriani}}, \bibinfo {author}
  {\bibfnamefont {S.}~\bibnamefont {Heinze}}, \bibinfo {author} {\bibfnamefont
  {G.}~\bibnamefont {Bihlmayer}}, \bibinfo {author} {\bibfnamefont
  {A.}~\bibnamefont {Kubetzka}}, \bibinfo {author} {\bibfnamefont
  {O.}~\bibnamefont {Pietzsch}}, \bibinfo {author} {\bibfnamefont
  {S.}~\bibnamefont {Bl{\"u}gel}}, \ and\ \bibinfo {author} {\bibfnamefont
  {R.}~\bibnamefont {Wiesendanger}},\ }\bibfield  {title} {\enquote {\bibinfo
  {title} {{Chiral magnetic order at surfaces driven by inversion
  asymmetry}},}\ }\href {\doibase 10.1038/nature05802} {\bibfield  {journal}
  {\bibinfo  {journal} {Nature}\ }\textbf {\bibinfo {volume} {447}},\ \bibinfo
  {pages} {190--193} (\bibinfo {year} {2007})}\BibitemShut {NoStop}%
\bibitem [{\citenamefont {Fern\'andez-Rossier}(2009)}]{Rossier2009_stm}%
  \BibitemOpen
  \bibfield  {author} {\bibinfo {author} {\bibfnamefont {J.}~\bibnamefont
  {Fern\'andez-Rossier}},\ }\bibfield  {title} {\enquote {\bibinfo {title}
  {{Theory of Single-Spin Inelastic Tunneling Spectroscopy}},}\ }\href
  {\doibase 10.1103/PhysRevLett.102.256802} {\bibfield  {journal} {\bibinfo
  {journal} {Phys. Rev. Lett.}\ }\textbf {\bibinfo {volume} {102}},\ \bibinfo
  {pages} {256802} (\bibinfo {year} {2009})}\BibitemShut {NoStop}%
\bibitem [{\citenamefont {Fransson}\ \emph {et~al.}(2010)\citenamefont
  {Fransson}, \citenamefont {Eriksson},\ and\ \citenamefont
  {Balatsky}}]{Balatsky2010_stm}%
  \BibitemOpen
  \bibfield  {author} {\bibinfo {author} {\bibfnamefont {J.}~\bibnamefont
  {Fransson}}, \bibinfo {author} {\bibfnamefont {O.}~\bibnamefont {Eriksson}},
  \ and\ \bibinfo {author} {\bibfnamefont {A.~V.}\ \bibnamefont {Balatsky}},\
  }\bibfield  {title} {\enquote {\bibinfo {title} {{Theory of spin-polarized
  scanning tunneling microscopy applied to local spins}},}\ }\href {\doibase
  10.1103/PhysRevB.81.115454} {\bibfield  {journal} {\bibinfo  {journal} {Phys.
  Rev. B}\ }\textbf {\bibinfo {volume} {81}},\ \bibinfo {pages} {115454}
  (\bibinfo {year} {2010})}\BibitemShut {NoStop}%
\bibitem [{\citenamefont {{Balashov, T. and Tak\'acs, A. F. and Wulfhekel, W.
  and Kirschner, J.}}(2006)}]{Balashov2006_magnon}%
  \BibitemOpen
  \bibfield  {author} {\bibinfo {author} {\bibnamefont {{Balashov, T. and
  Tak\'acs, A. F. and Wulfhekel, W. and Kirschner, J.}}},\ }\bibfield  {title}
  {\enquote {\bibinfo {title} {{Magnon Excitation with Spin-Polarized Scanning
  Tunneling Microscopy}},}\ }\href {\doibase 10.1103/PhysRevLett.97.187201}
  {\bibfield  {journal} {\bibinfo  {journal} {Phys. Rev. Lett.}\ }\textbf
  {\bibinfo {volume} {97}},\ \bibinfo {pages} {187201} (\bibinfo {year}
  {2006})}\BibitemShut {NoStop}%
\bibitem [{\citenamefont {Spinelli}\ \emph {et~al.}(2014)\citenamefont
  {Spinelli}, \citenamefont {Bryant}, \citenamefont {Delgado}, \citenamefont
  {Fern{\'a}ndez-Rossier},\ and\ \citenamefont {Otte}}]{Spinelli2014_sw}%
  \BibitemOpen
  \bibfield  {author} {\bibinfo {author} {\bibfnamefont {A.}~\bibnamefont
  {Spinelli}}, \bibinfo {author} {\bibfnamefont {B.}~\bibnamefont {Bryant}},
  \bibinfo {author} {\bibfnamefont {F.}~\bibnamefont {Delgado}}, \bibinfo
  {author} {\bibfnamefont {J.}~\bibnamefont {Fern{\'a}ndez-Rossier}}, \ and\
  \bibinfo {author} {\bibfnamefont {A.~F.}\ \bibnamefont {Otte}},\ }\bibfield
  {title} {\enquote {\bibinfo {title} {{Imaging of spin waves in atomically
  designed nanomagnets}},}\ }\href {\doibase 10.1038/nmat4018} {\bibfield
  {journal} {\bibinfo  {journal} {Nature Materials}\ }\textbf {\bibinfo
  {volume} {13}},\ \bibinfo {pages} {782--785} (\bibinfo {year}
  {2014})}\BibitemShut {NoStop}%
\bibitem [{\citenamefont {Delgado}\ \emph {et~al.}(2013)\citenamefont
  {Delgado}, \citenamefont {Batista},\ and\ \citenamefont
  {Fern\'andez-Rossier}}]{Delgado2013_fract}%
  \BibitemOpen
  \bibfield  {author} {\bibinfo {author} {\bibfnamefont {F.}~\bibnamefont
  {Delgado}}, \bibinfo {author} {\bibfnamefont {C.~D.}\ \bibnamefont
  {Batista}}, \ and\ \bibinfo {author} {\bibfnamefont {J.}~\bibnamefont
  {Fern\'andez-Rossier}},\ }\bibfield  {title} {\enquote {\bibinfo {title}
  {{Local Probe of Fractional Edge States of $S=1$ Heisenberg Spin Chains}},}\
  }\href {\doibase 10.1103/PhysRevLett.111.167201} {\bibfield  {journal}
  {\bibinfo  {journal} {Phys. Rev. Lett.}\ }\textbf {\bibinfo {volume} {111}},\
  \bibinfo {pages} {167201} (\bibinfo {year} {2013})}\BibitemShut {NoStop}%
\bibitem [{\citenamefont {Onose}\ \emph {et~al.}(2010)\citenamefont {Onose},
  \citenamefont {Ideue}, \citenamefont {Katsura}, \citenamefont {Shiomi},
  \citenamefont {Nagaosa},\ and\ \citenamefont
  {Tokura}}]{onose2010observation}%
  \BibitemOpen
  \bibfield  {author} {\bibinfo {author} {\bibfnamefont {Y.}~\bibnamefont
  {Onose}}, \bibinfo {author} {\bibfnamefont {T.}~\bibnamefont {Ideue}},
  \bibinfo {author} {\bibfnamefont {H.}~\bibnamefont {Katsura}}, \bibinfo
  {author} {\bibfnamefont {Y.}~\bibnamefont {Shiomi}}, \bibinfo {author}
  {\bibfnamefont {N.}~\bibnamefont {Nagaosa}}, \ and\ \bibinfo {author}
  {\bibfnamefont {Y.}~\bibnamefont {Tokura}},\ }\bibfield  {title} {\enquote
  {\bibinfo {title} {{Observation of the Magnon Hall Effect}},}\ }\href
  {\doibase 10.1126/science.1188260} {\bibfield  {journal} {\bibinfo  {journal}
  {Science}\ }\textbf {\bibinfo {volume} {329}},\ \bibinfo {pages} {297--299}
  (\bibinfo {year} {2010})}\BibitemShut {NoStop}%
\bibitem [{\citenamefont {Hirschberger}\ \emph {et~al.}(2015)\citenamefont
  {Hirschberger}, \citenamefont {Chisnell}, \citenamefont {Lee},\ and\
  \citenamefont {Ong}}]{hirschberger2015thermal}%
  \BibitemOpen
  \bibfield  {author} {\bibinfo {author} {\bibfnamefont {M.}~\bibnamefont
  {Hirschberger}}, \bibinfo {author} {\bibfnamefont {R.}~\bibnamefont
  {Chisnell}}, \bibinfo {author} {\bibfnamefont {Y.~S.}\ \bibnamefont {Lee}}, \
  and\ \bibinfo {author} {\bibfnamefont {N.~P.}\ \bibnamefont {Ong}},\
  }\bibfield  {title} {\enquote {\bibinfo {title} {{Thermal Hall Effect of Spin
  Excitations in a Kagome Magnet}},}\ }\href {\doibase
  10.1103/PhysRevLett.115.106603} {\bibfield  {journal} {\bibinfo  {journal}
  {Phys. Rev. Lett.}\ }\textbf {\bibinfo {volume} {115}},\ \bibinfo {pages}
  {106603} (\bibinfo {year} {2015})}\BibitemShut {NoStop}%
\bibitem [{\citenamefont {Chisnell}\ \emph {et~al.}(2015)\citenamefont
  {Chisnell}, \citenamefont {Helton}, \citenamefont {Freedman}, \citenamefont
  {Singh}, \citenamefont {Bewley}, \citenamefont {Nocera},\ and\ \citenamefont
  {Lee}}]{chisnell2015topological}%
  \BibitemOpen
  \bibfield  {author} {\bibinfo {author} {\bibfnamefont {R.}~\bibnamefont
  {Chisnell}}, \bibinfo {author} {\bibfnamefont {J.~S.}\ \bibnamefont
  {Helton}}, \bibinfo {author} {\bibfnamefont {D.~E.}\ \bibnamefont
  {Freedman}}, \bibinfo {author} {\bibfnamefont {D.~K.}\ \bibnamefont {Singh}},
  \bibinfo {author} {\bibfnamefont {R.~I.}\ \bibnamefont {Bewley}}, \bibinfo
  {author} {\bibfnamefont {D.~G.}\ \bibnamefont {Nocera}}, \ and\ \bibinfo
  {author} {\bibfnamefont {Y.~S.}\ \bibnamefont {Lee}},\ }\bibfield  {title}
  {\enquote {\bibinfo {title} {{Topological Magnon Bands in a Kagome Lattice
  Ferromagnet}},}\ }\href {\doibase 10.1103/PhysRevLett.115.147201} {\bibfield
  {journal} {\bibinfo  {journal} {Phys. Rev. Lett.}\ }\textbf {\bibinfo
  {volume} {115}},\ \bibinfo {pages} {147201} (\bibinfo {year}
  {2015})}\BibitemShut {NoStop}%
\bibitem [{\citenamefont {Rold{\'a}n-Molina}\ \emph {et~al.}(2016)\citenamefont
  {Rold{\'a}n-Molina}, \citenamefont {Nunez},\ and\ \citenamefont
  {Fern{\'a}ndez-Rossier}}]{Molina2016_topo}%
  \BibitemOpen
  \bibfield  {author} {\bibinfo {author} {\bibfnamefont {A.}~\bibnamefont
  {Rold{\'a}n-Molina}}, \bibinfo {author} {\bibfnamefont {A.~S.}\ \bibnamefont
  {Nunez}}, \ and\ \bibinfo {author} {\bibfnamefont {J.}~\bibnamefont
  {Fern{\'a}ndez-Rossier}},\ }\bibfield  {title} {\enquote {\bibinfo {title}
  {{Topological spin waves in the atomic-scale magnetic skyrmion crystal}},}\
  }\href {\doibase 10.1088/1367-2630/18/4/045015} {\bibfield  {journal}
  {\bibinfo  {journal} {New Journal of Physics}\ }\textbf {\bibinfo {volume}
  {18}},\ \bibinfo {pages} {045015} (\bibinfo {year} {2016})}\BibitemShut
  {NoStop}%
\bibitem [{\citenamefont {Huang}\ \emph {et~al.}(2017)\citenamefont {Huang},
  \citenamefont {Clark}, \citenamefont {Navarro-Moratalla}, \citenamefont
  {Klein}, \citenamefont {Cheng}, \citenamefont {Seyler}, \citenamefont
  {Zhong}, \citenamefont {Schmidgall}, \citenamefont {McGuire}, \citenamefont
  {Cobden},\ and\ \citenamefont {et~al.}}]{Huang2017_layer}%
  \BibitemOpen
  \bibfield  {author} {\bibinfo {author} {\bibfnamefont {B.}~\bibnamefont
  {Huang}}, \bibinfo {author} {\bibfnamefont {G.}~\bibnamefont {Clark}},
  \bibinfo {author} {\bibfnamefont {E.}~\bibnamefont {Navarro-Moratalla}},
  \bibinfo {author} {\bibfnamefont {D.~R.}\ \bibnamefont {Klein}}, \bibinfo
  {author} {\bibfnamefont {R.}~\bibnamefont {Cheng}}, \bibinfo {author}
  {\bibfnamefont {K.~L.}\ \bibnamefont {Seyler}}, \bibinfo {author}
  {\bibfnamefont {D.}~\bibnamefont {Zhong}}, \bibinfo {author} {\bibfnamefont
  {E.}~\bibnamefont {Schmidgall}}, \bibinfo {author} {\bibfnamefont {M.~A.}\
  \bibnamefont {McGuire}}, \bibinfo {author} {\bibfnamefont {D.~H.}\
  \bibnamefont {Cobden}}, \ and\ \bibinfo {author} {\bibnamefont {et~al.}},\
  }\bibfield  {title} {\enquote {\bibinfo {title} {{Layer-dependent
  ferromagnetism in a van der Waals crystal down to the monolayer limit}},}\
  }\href {\doibase 10.1038/nature22391} {\bibfield  {journal} {\bibinfo
  {journal} {Nature}\ }\textbf {\bibinfo {volume} {546}},\ \bibinfo {pages}
  {270--273} (\bibinfo {year} {2017})}\BibitemShut {NoStop}%
\bibitem [{\citenamefont {Aguilera}\ \emph {et~al.}(2020)\citenamefont
  {Aguilera}, \citenamefont {Jaeschke-Ubiergo}, \citenamefont {Vidal-Silva},
  \citenamefont {Foa~Torres},\ and\ \citenamefont
  {Nunez}}]{Aguilera2020_magnon}%
  \BibitemOpen
  \bibfield  {author} {\bibinfo {author} {\bibfnamefont {E.}~\bibnamefont
  {Aguilera}}, \bibinfo {author} {\bibfnamefont {R.}~\bibnamefont
  {Jaeschke-Ubiergo}}, \bibinfo {author} {\bibfnamefont {N.}~\bibnamefont
  {Vidal-Silva}}, \bibinfo {author} {\bibfnamefont {L.E.F}\ \bibnamefont
  {Foa~Torres}}, \ and\ \bibinfo {author} {\bibfnamefont {A.S.}\ \bibnamefont
  {Nunez}},\ }\href@noop {} {\enquote {\bibinfo {title} {{Topological magnonics
  in the two-dimensional van der Waals magnet CrI3}},}\ } (\bibinfo {year}
  {2020}),\ \Eprint {http://arxiv.org/abs/2002.05266} {arXiv:2002.05266}
  \BibitemShut {NoStop}%
\bibitem [{\citenamefont {{A. Kitaev}}(2006)}]{kitaev_2006}%
  \BibitemOpen
  \bibfield  {author} {\bibinfo {author} {\bibnamefont {{A. Kitaev}}},\
  }\bibfield  {title} {\enquote {\bibinfo {title} {{Anyons in an exactly solved
  model and beyond}},}\ }\href {\doibase
  https://doi.org/10.1016/j.aop.2005.10.005} {\bibfield  {journal} {\bibinfo
  {journal} {Annals of Physics}\ }\textbf {\bibinfo {volume} {321}},\ \bibinfo
  {pages} {2 -- 111} (\bibinfo {year} {2006})},\ \bibinfo {note} {january
  Special Issue}\BibitemShut {NoStop}%
\bibitem [{\citenamefont {Jackeli}\ and\ \citenamefont
  {Khaliullin}(2009)}]{Jackeli2009_mott}%
  \BibitemOpen
  \bibfield  {author} {\bibinfo {author} {\bibfnamefont {G.}~\bibnamefont
  {Jackeli}}\ and\ \bibinfo {author} {\bibfnamefont {G.}~\bibnamefont
  {Khaliullin}},\ }\bibfield  {title} {\enquote {\bibinfo {title} {{Mott
  Insulators in the Strong Spin-Orbit Coupling Limit: From Heisenberg to a
  Quantum Compass and Kitaev Models}},}\ }\href {\doibase
  10.1103/PhysRevLett.102.017205} {\bibfield  {journal} {\bibinfo  {journal}
  {Phys. Rev. Lett.}\ }\textbf {\bibinfo {volume} {102}},\ \bibinfo {pages}
  {017205} (\bibinfo {year} {2009})}\BibitemShut {NoStop}%
\bibitem [{\citenamefont {Winter}\ \emph {et~al.}(2017)\citenamefont {Winter},
  \citenamefont {Tsirlin}, \citenamefont {Daghofer}, \citenamefont {van~den
  Brink}, \citenamefont {Singh}, \citenamefont {Gegenwart},\ and\ \citenamefont
  {Valent{\'\i}}}]{Winter2017_models}%
  \BibitemOpen
  \bibfield  {author} {\bibinfo {author} {\bibfnamefont {S.~M.}\ \bibnamefont
  {Winter}}, \bibinfo {author} {\bibfnamefont {A.~A.}\ \bibnamefont {Tsirlin}},
  \bibinfo {author} {\bibfnamefont {M.}~\bibnamefont {Daghofer}}, \bibinfo
  {author} {\bibfnamefont {J.}~\bibnamefont {van~den Brink}}, \bibinfo {author}
  {\bibfnamefont {Y.}~\bibnamefont {Singh}}, \bibinfo {author} {\bibfnamefont
  {P.}~\bibnamefont {Gegenwart}}, \ and\ \bibinfo {author} {\bibfnamefont
  {R.}~\bibnamefont {Valent{\'\i}}},\ }\bibfield  {title} {\enquote {\bibinfo
  {title} {{Models and materials for generalized Kitaev magnetism}},}\ }\href
  {\doibase 10.1088/1361-648x/aa8cf5} {\bibfield  {journal} {\bibinfo
  {journal} {Journal of Physics: Condensed Matter}\ }\textbf {\bibinfo {volume}
  {29}},\ \bibinfo {pages} {493002} (\bibinfo {year} {2017})}\BibitemShut
  {NoStop}%
\bibitem [{\citenamefont {Hermanns}\ \emph {et~al.}(2018)\citenamefont
  {Hermanns}, \citenamefont {Kimchi},\ and\ \citenamefont
  {Knolle}}]{Hermanns2018_physics}%
  \BibitemOpen
  \bibfield  {author} {\bibinfo {author} {\bibfnamefont {M.}~\bibnamefont
  {Hermanns}}, \bibinfo {author} {\bibfnamefont {I.}~\bibnamefont {Kimchi}}, \
  and\ \bibinfo {author} {\bibfnamefont {J.}~\bibnamefont {Knolle}},\
  }\bibfield  {title} {\enquote {\bibinfo {title} {{Physics of the Kitaev
  Model: Fractionalization, Dynamic Correlations, and Material Connections}},}\
  }\href {\doibase 10.1146/annurev-conmatphys-033117-053934} {\bibfield
  {journal} {\bibinfo  {journal} {Annual Review of Condensed Matter Physics}\
  }\textbf {\bibinfo {volume} {9}},\ \bibinfo {pages} {17--33} (\bibinfo {year}
  {2018})}\BibitemShut {NoStop}%
\bibitem [{\citenamefont {Takagi}\ \emph {et~al.}(2019)\citenamefont {Takagi},
  \citenamefont {Takayama}, \citenamefont {Jackeli}, \citenamefont
  {Khaliullin},\ and\ \citenamefont {Nagler}}]{Takagi2019_concept}%
  \BibitemOpen
  \bibfield  {author} {\bibinfo {author} {\bibfnamefont {H.}~\bibnamefont
  {Takagi}}, \bibinfo {author} {\bibfnamefont {T.}~\bibnamefont {Takayama}},
  \bibinfo {author} {\bibfnamefont {G.}~\bibnamefont {Jackeli}}, \bibinfo
  {author} {\bibfnamefont {G.}~\bibnamefont {Khaliullin}}, \ and\ \bibinfo
  {author} {\bibfnamefont {S.~E.}\ \bibnamefont {Nagler}},\ }\bibfield  {title}
  {\enquote {\bibinfo {title} {{Concept and realization of Kitaev quantum spin
  liquids}},}\ }\href {\doibase 10.1038/s42254-019-0038-2} {\bibfield
  {journal} {\bibinfo  {journal} {Nature Reviews Physics}\ }\textbf {\bibinfo
  {volume} {1}},\ \bibinfo {pages} {264--280} (\bibinfo {year}
  {2019})}\BibitemShut {NoStop}%
\bibitem [{\citenamefont {Zhou}\ \emph
  {et~al.}(2019{\natexlab{a}})\citenamefont {Zhou}, \citenamefont {Wang},
  \citenamefont {Osterhoudt}, \citenamefont {Lampen-Kelley}, \citenamefont
  {Mandrus}, \citenamefont {He}, \citenamefont {Burch},\ and\ \citenamefont
  {Henriksen}}]{zhou2019possible}%
  \BibitemOpen
  \bibfield  {author} {\bibinfo {author} {\bibfnamefont {B.}~\bibnamefont
  {Zhou}}, \bibinfo {author} {\bibfnamefont {Y.}~\bibnamefont {Wang}}, \bibinfo
  {author} {\bibfnamefont {G.~B.}\ \bibnamefont {Osterhoudt}}, \bibinfo
  {author} {\bibfnamefont {P.}~\bibnamefont {Lampen-Kelley}}, \bibinfo {author}
  {\bibfnamefont {D.}~\bibnamefont {Mandrus}}, \bibinfo {author} {\bibfnamefont
  {R.}~\bibnamefont {He}}, \bibinfo {author} {\bibfnamefont {K.~S.}\
  \bibnamefont {Burch}}, \ and\ \bibinfo {author} {\bibfnamefont {E.~A.}\
  \bibnamefont {Henriksen}},\ }\bibfield  {title} {\enquote {\bibinfo {title}
  {{Possible structural transformation and enhanced magnetic fluctuations in
  exfoliated $\alpha$-RuCl3}},}\ }\href {\doibase
  https://doi.org/10.1016/j.jpcs.2018.01.026} {\bibfield  {journal} {\bibinfo
  {journal} {Journal of Physics and Chemistry of Solids}\ }\textbf {\bibinfo
  {volume} {128}},\ \bibinfo {pages} {291 -- 295} (\bibinfo {year}
  {2019}{\natexlab{a}})},\ \bibinfo {note} {spin-Orbit Coupled
  Materials}\BibitemShut {NoStop}%
\bibitem [{\citenamefont {Zhou}\ \emph
  {et~al.}(2019{\natexlab{b}})\citenamefont {Zhou}, \citenamefont {Balgley},
  \citenamefont {Lampen-Kelley}, \citenamefont {Yan}, \citenamefont {Mandrus},\
  and\ \citenamefont {Henriksen}}]{PhysRevB.100.165426}%
  \BibitemOpen
  \bibfield  {author} {\bibinfo {author} {\bibfnamefont {B.}~\bibnamefont
  {Zhou}}, \bibinfo {author} {\bibfnamefont {J.}~\bibnamefont {Balgley}},
  \bibinfo {author} {\bibfnamefont {P.}~\bibnamefont {Lampen-Kelley}}, \bibinfo
  {author} {\bibfnamefont {J.-Q.}\ \bibnamefont {Yan}}, \bibinfo {author}
  {\bibfnamefont {D.~G.}\ \bibnamefont {Mandrus}}, \ and\ \bibinfo {author}
  {\bibfnamefont {E.~A.}\ \bibnamefont {Henriksen}},\ }\bibfield  {title}
  {\enquote {\bibinfo {title} {{Evidence for charge transfer and proximate
  magnetism in
  graphene--$\ensuremath{\alpha}\text{\ensuremath{-}}{\mathrm{RuCl}}_{3}$
  heterostructures}},}\ }\href {\doibase 10.1103/PhysRevB.100.165426}
  {\bibfield  {journal} {\bibinfo  {journal} {Phys. Rev. B}\ }\textbf {\bibinfo
  {volume} {100}},\ \bibinfo {pages} {165426} (\bibinfo {year}
  {2019}{\natexlab{b}})}\BibitemShut {NoStop}%
\bibitem [{\citenamefont {Mashhadi}\ \emph {et~al.}(2019)\citenamefont
  {Mashhadi}, \citenamefont {Kim}, \citenamefont {Kim}, \citenamefont {Weber},
  \citenamefont {Taniguchi}, \citenamefont {Watanabe}, \citenamefont {Park},
  \citenamefont {Lotsch}, \citenamefont {Smet}, \citenamefont {Burghard},\ and\
  \citenamefont {et~al.}}]{mashhadi2019spin}%
  \BibitemOpen
  \bibfield  {author} {\bibinfo {author} {\bibfnamefont {S.}~\bibnamefont
  {Mashhadi}}, \bibinfo {author} {\bibfnamefont {Y.}~\bibnamefont {Kim}},
  \bibinfo {author} {\bibfnamefont {J.}~\bibnamefont {Kim}}, \bibinfo {author}
  {\bibfnamefont {D.}~\bibnamefont {Weber}}, \bibinfo {author} {\bibfnamefont
  {T.}~\bibnamefont {Taniguchi}}, \bibinfo {author} {\bibfnamefont
  {K.}~\bibnamefont {Watanabe}}, \bibinfo {author} {\bibfnamefont
  {N.}~\bibnamefont {Park}}, \bibinfo {author} {\bibfnamefont {B.}~\bibnamefont
  {Lotsch}}, \bibinfo {author} {\bibfnamefont {J.~H.}\ \bibnamefont {Smet}},
  \bibinfo {author} {\bibfnamefont {M.}~\bibnamefont {Burghard}}, \ and\
  \bibinfo {author} {\bibnamefont {et~al.}},\ }\bibfield  {title} {\enquote
  {\bibinfo {title} {{Spin-Split Band Hybridization in Graphene Proximitized
  with $\alpha$-RuCl3 Nanosheets}},}\ }\href {\doibase
  10.1021/acs.nanolett.9b01691} {\bibfield  {journal} {\bibinfo  {journal}
  {Nano Letters}\ }\textbf {\bibinfo {volume} {19}},\ \bibinfo {pages}
  {4659--4665} (\bibinfo {year} {2019})}\BibitemShut {NoStop}%
\bibitem [{\citenamefont {Sears}\ \emph {et~al.}(2015)\citenamefont {Sears},
  \citenamefont {Songvilay}, \citenamefont {Plumb}, \citenamefont {Clancy},
  \citenamefont {Qiu}, \citenamefont {Zhao}, \citenamefont {Parshall},\ and\
  \citenamefont {Kim}}]{Sears2015_magnetic}%
  \BibitemOpen
  \bibfield  {author} {\bibinfo {author} {\bibfnamefont {J.~A.}\ \bibnamefont
  {Sears}}, \bibinfo {author} {\bibfnamefont {M.}~\bibnamefont {Songvilay}},
  \bibinfo {author} {\bibfnamefont {K.~W.}\ \bibnamefont {Plumb}}, \bibinfo
  {author} {\bibfnamefont {J.~P.}\ \bibnamefont {Clancy}}, \bibinfo {author}
  {\bibfnamefont {Y.}~\bibnamefont {Qiu}}, \bibinfo {author} {\bibfnamefont
  {Y.}~\bibnamefont {Zhao}}, \bibinfo {author} {\bibfnamefont {D.}~\bibnamefont
  {Parshall}}, \ and\ \bibinfo {author} {\bibfnamefont {Young-June}\
  \bibnamefont {Kim}},\ }\bibfield  {title} {\enquote {\bibinfo {title}
  {{Magnetic order in $\ensuremath{\alpha}\ensuremath{-}{\text{RuCl}}_{3}$: A
  honeycomb-lattice quantum magnet with strong spin-orbit coupling}},}\ }\href
  {\doibase 10.1103/PhysRevB.91.144420} {\bibfield  {journal} {\bibinfo
  {journal} {Phys. Rev. B}\ }\textbf {\bibinfo {volume} {91}},\ \bibinfo
  {pages} {144420} (\bibinfo {year} {2015})}\BibitemShut {NoStop}%
\bibitem [{\citenamefont {Johnson}\ \emph {et~al.}(2015)\citenamefont
  {Johnson}, \citenamefont {Williams}, \citenamefont {Haghighirad},
  \citenamefont {Singleton}, \citenamefont {Zapf}, \citenamefont {Manuel},
  \citenamefont {Mazin}, \citenamefont {Li}, \citenamefont {Jeschke},
  \citenamefont {Valent\'{\i}},\ and\ \citenamefont
  {Coldea}}]{Johnson2015_monoclinic}%
  \BibitemOpen
  \bibfield  {author} {\bibinfo {author} {\bibfnamefont {R.~D.}\ \bibnamefont
  {Johnson}}, \bibinfo {author} {\bibfnamefont {S.~C.}\ \bibnamefont
  {Williams}}, \bibinfo {author} {\bibfnamefont {A.~A.}\ \bibnamefont
  {Haghighirad}}, \bibinfo {author} {\bibfnamefont {J.}~\bibnamefont
  {Singleton}}, \bibinfo {author} {\bibfnamefont {V.}~\bibnamefont {Zapf}},
  \bibinfo {author} {\bibfnamefont {P.}~\bibnamefont {Manuel}}, \bibinfo
  {author} {\bibfnamefont {I.~I.}\ \bibnamefont {Mazin}}, \bibinfo {author}
  {\bibfnamefont {Y.}~\bibnamefont {Li}}, \bibinfo {author} {\bibfnamefont
  {H.~O.}\ \bibnamefont {Jeschke}}, \bibinfo {author} {\bibfnamefont
  {R.}~\bibnamefont {Valent\'{\i}}}, \ and\ \bibinfo {author} {\bibfnamefont
  {R.}~\bibnamefont {Coldea}},\ }\bibfield  {title} {\enquote {\bibinfo {title}
  {{Monoclinic crystal structure of
  $\ensuremath{\alpha}\ensuremath{-}{\mathrm{RuCl}}_{3}$ and the zigzag
  antiferromagnetic ground state}},}\ }\href {\doibase
  10.1103/PhysRevB.92.235119} {\bibfield  {journal} {\bibinfo  {journal} {Phys.
  Rev. B}\ }\textbf {\bibinfo {volume} {92}},\ \bibinfo {pages} {235119}
  (\bibinfo {year} {2015})}\BibitemShut {NoStop}%
\bibitem [{\citenamefont {Banerjee}\ \emph {et~al.}(2016)\citenamefont
  {Banerjee}, \citenamefont {Bridges}, \citenamefont {Yan}, \citenamefont
  {Aczel}, \citenamefont {Li}, \citenamefont {Stone}, \citenamefont {Granroth},
  \citenamefont {Lumsden}, \citenamefont {Yiu}, \citenamefont {Knolle},\ and\
  \citenamefont {et~al.}}]{Banerjee2016_proximate}%
  \BibitemOpen
  \bibfield  {author} {\bibinfo {author} {\bibfnamefont {A.}~\bibnamefont
  {Banerjee}}, \bibinfo {author} {\bibfnamefont {C.~A.}\ \bibnamefont
  {Bridges}}, \bibinfo {author} {\bibfnamefont {J.-Q.}\ \bibnamefont {Yan}},
  \bibinfo {author} {\bibfnamefont {A.~A.}\ \bibnamefont {Aczel}}, \bibinfo
  {author} {\bibfnamefont {L.}~\bibnamefont {Li}}, \bibinfo {author}
  {\bibfnamefont {M.~B.}\ \bibnamefont {Stone}}, \bibinfo {author}
  {\bibfnamefont {G.~E.}\ \bibnamefont {Granroth}}, \bibinfo {author}
  {\bibfnamefont {M.~D.}\ \bibnamefont {Lumsden}}, \bibinfo {author}
  {\bibfnamefont {Y.}~\bibnamefont {Yiu}}, \bibinfo {author} {\bibfnamefont
  {J.}~\bibnamefont {Knolle}}, \ and\ \bibinfo {author} {\bibnamefont
  {et~al.}},\ }\bibfield  {title} {\enquote {\bibinfo {title} {{Proximate
  Kitaev quantum spin liquid behaviour in a honeycomb magnet}},}\ }\href
  {\doibase 10.1038/nmat4604} {\bibfield  {journal} {\bibinfo  {journal}
  {Nature Materials}\ }\textbf {\bibinfo {volume} {15}},\ \bibinfo {pages}
  {733--740} (\bibinfo {year} {2016})}\BibitemShut {NoStop}%
\bibitem [{\citenamefont {Banerjee}\ \emph {et~al.}(2017)\citenamefont
  {Banerjee}, \citenamefont {Yan}, \citenamefont {Knolle}, \citenamefont
  {Bridges}, \citenamefont {Stone}, \citenamefont {Lumsden}, \citenamefont
  {Mandrus}, \citenamefont {Tennant}, \citenamefont {Moessner},\ and\
  \citenamefont {Nagler}}]{Banerjee2017_neutron}%
  \BibitemOpen
  \bibfield  {author} {\bibinfo {author} {\bibfnamefont {A.}~\bibnamefont
  {Banerjee}}, \bibinfo {author} {\bibfnamefont {J.}~\bibnamefont {Yan}},
  \bibinfo {author} {\bibfnamefont {J.}~\bibnamefont {Knolle}}, \bibinfo
  {author} {\bibfnamefont {C.~A.}\ \bibnamefont {Bridges}}, \bibinfo {author}
  {\bibfnamefont {M.~B.}\ \bibnamefont {Stone}}, \bibinfo {author}
  {\bibfnamefont {M.~D.}\ \bibnamefont {Lumsden}}, \bibinfo {author}
  {\bibfnamefont {D.~G.}\ \bibnamefont {Mandrus}}, \bibinfo {author}
  {\bibfnamefont {D.~A.}\ \bibnamefont {Tennant}}, \bibinfo {author}
  {\bibfnamefont {R.}~\bibnamefont {Moessner}}, \ and\ \bibinfo {author}
  {\bibfnamefont {S.~E.}\ \bibnamefont {Nagler}},\ }\bibfield  {title}
  {\enquote {\bibinfo {title} {{Neutron scattering in the proximate quantum
  spin liquid $\alpha$-RuCl3}},}\ }\href {\doibase 10.1126/science.aah6015}
  {\bibfield  {journal} {\bibinfo  {journal} {Science}\ }\textbf {\bibinfo
  {volume} {356}},\ \bibinfo {pages} {1055--1059} (\bibinfo {year}
  {2017})}\BibitemShut {NoStop}%
\bibitem [{\citenamefont {Banerjee}\ \emph {et~al.}(2018)\citenamefont
  {Banerjee}, \citenamefont {Lampen-Kelley}, \citenamefont {Knolle},
  \citenamefont {Balz}, \citenamefont {Aczel}, \citenamefont {Winn},
  \citenamefont {Liu}, \citenamefont {Pajerowski}, \citenamefont {Yan},
  \citenamefont {Bridges},\ and\ \citenamefont
  {et~al.}}]{Banerjee2018_excitations}%
  \BibitemOpen
  \bibfield  {author} {\bibinfo {author} {\bibfnamefont {A.}~\bibnamefont
  {Banerjee}}, \bibinfo {author} {\bibfnamefont {P.}~\bibnamefont
  {Lampen-Kelley}}, \bibinfo {author} {\bibfnamefont {J.}~\bibnamefont
  {Knolle}}, \bibinfo {author} {\bibfnamefont {C.}~\bibnamefont {Balz}},
  \bibinfo {author} {\bibfnamefont {A.~A.}\ \bibnamefont {Aczel}}, \bibinfo
  {author} {\bibfnamefont {B.}~\bibnamefont {Winn}}, \bibinfo {author}
  {\bibfnamefont {Y.}~\bibnamefont {Liu}}, \bibinfo {author} {\bibfnamefont
  {D.}~\bibnamefont {Pajerowski}}, \bibinfo {author} {\bibfnamefont
  {J.}~\bibnamefont {Yan}}, \bibinfo {author} {\bibfnamefont {C.~A.}\
  \bibnamefont {Bridges}}, \ and\ \bibinfo {author} {\bibnamefont {et~al.}},\
  }\bibfield  {title} {\enquote {\bibinfo {title} {{Excitations in the
  field-induced quantum spin liquid state of $\alpha-\mathrm{RuCl}_3$}},}\
  }\href {\doibase 10.1038/s41535-018-0079-2} {\bibfield  {journal} {\bibinfo
  {journal} {npj Quantum Materials}\ }\textbf {\bibinfo {volume} {3}} (\bibinfo
  {year} {2018}),\ 10.1038/s41535-018-0079-2}\BibitemShut {NoStop}%
\bibitem [{\citenamefont {Winter}\ \emph {et~al.}(2018)\citenamefont {Winter},
  \citenamefont {Riedl}, \citenamefont {Kaib}, \citenamefont {Coldea},\ and\
  \citenamefont {Valent\'{\i}}}]{Winter2018_probing}%
  \BibitemOpen
  \bibfield  {author} {\bibinfo {author} {\bibfnamefont {S.~M.}\ \bibnamefont
  {Winter}}, \bibinfo {author} {\bibfnamefont {K.}~\bibnamefont {Riedl}},
  \bibinfo {author} {\bibfnamefont {D.}~\bibnamefont {Kaib}}, \bibinfo {author}
  {\bibfnamefont {R.}~\bibnamefont {Coldea}}, \ and\ \bibinfo {author}
  {\bibfnamefont {R.}~\bibnamefont {Valent\'{\i}}},\ }\bibfield  {title}
  {\enquote {\bibinfo {title} {{Probing
  $\ensuremath{\alpha}\ensuremath{-}{\mathrm{RuCl}}_{3}$ Beyond Magnetic Order:
  Effects of Temperature and Magnetic Field}},}\ }\href {\doibase
  10.1103/PhysRevLett.120.077203} {\bibfield  {journal} {\bibinfo  {journal}
  {Phys. Rev. Lett.}\ }\textbf {\bibinfo {volume} {120}},\ \bibinfo {pages}
  {077203} (\bibinfo {year} {2018})}\BibitemShut {NoStop}%
\bibitem [{\citenamefont {Kasahara}\ \emph {et~al.}(2018)\citenamefont
  {Kasahara}, \citenamefont {Ohnishi}, \citenamefont {Mizukami}, \citenamefont
  {Tanaka}, \citenamefont {Ma}, \citenamefont {Sugii}, \citenamefont {Kurita},
  \citenamefont {Tanaka}, \citenamefont {Nasu}, \citenamefont {Motome},\ and\
  \citenamefont {et~al.}}]{Kasahara2018_majorana}%
  \BibitemOpen
  \bibfield  {author} {\bibinfo {author} {\bibfnamefont {Y.}~\bibnamefont
  {Kasahara}}, \bibinfo {author} {\bibfnamefont {T.}~\bibnamefont {Ohnishi}},
  \bibinfo {author} {\bibfnamefont {Y.}~\bibnamefont {Mizukami}}, \bibinfo
  {author} {\bibfnamefont {O.}~\bibnamefont {Tanaka}}, \bibinfo {author}
  {\bibfnamefont {Sixiao}\ \bibnamefont {Ma}}, \bibinfo {author} {\bibfnamefont
  {K.}~\bibnamefont {Sugii}}, \bibinfo {author} {\bibfnamefont
  {N.}~\bibnamefont {Kurita}}, \bibinfo {author} {\bibfnamefont
  {H.}~\bibnamefont {Tanaka}}, \bibinfo {author} {\bibfnamefont
  {J.}~\bibnamefont {Nasu}}, \bibinfo {author} {\bibfnamefont {Y.}~\bibnamefont
  {Motome}}, \ and\ \bibinfo {author} {\bibnamefont {et~al.}},\ }\bibfield
  {title} {\enquote {\bibinfo {title} {{Majorana quantization and half-integer
  thermal quantum Hall effect in a Kitaev spin liquid}},}\ }\href {\doibase
  10.1038/s41586-018-0274-0} {\bibfield  {journal} {\bibinfo  {journal}
  {Nature}\ }\textbf {\bibinfo {volume} {559}},\ \bibinfo {pages} {227--231}
  (\bibinfo {year} {2018})}\BibitemShut {NoStop}%
\bibitem [{\citenamefont {Vinkler-Aviv}\ and\ \citenamefont
  {Rosch}(2018)}]{vinkler2018approximately}%
  \BibitemOpen
  \bibfield  {author} {\bibinfo {author} {\bibfnamefont {Y.}~\bibnamefont
  {Vinkler-Aviv}}\ and\ \bibinfo {author} {\bibfnamefont {A.}~\bibnamefont
  {Rosch}},\ }\bibfield  {title} {\enquote {\bibinfo {title} {{Approximately
  Quantized Thermal Hall Effect of Chiral Liquids Coupled to Phonons}},}\
  }\href {\doibase 10.1103/PhysRevX.8.031032} {\bibfield  {journal} {\bibinfo
  {journal} {Phys. Rev. X}\ }\textbf {\bibinfo {volume} {8}},\ \bibinfo {pages}
  {031032} (\bibinfo {year} {2018})}\BibitemShut {NoStop}%
\bibitem [{\citenamefont {{Ye, M. and Hal\'asz, G. B. and Savary, L. and
  Balents, L.}}(2018)}]{ye2018quantization}%
  \BibitemOpen
  \bibfield  {author} {\bibinfo {author} {\bibnamefont {{Ye, M. and Hal\'asz,
  G. B. and Savary, L. and Balents, L.}}},\ }\bibfield  {title} {\enquote
  {\bibinfo {title} {{Quantization of the Thermal Hall Conductivity at Small
  Hall Angles}},}\ }\href {\doibase 10.1103/PhysRevLett.121.147201} {\bibfield
  {journal} {\bibinfo  {journal} {Phys. Rev. Lett.}\ }\textbf {\bibinfo
  {volume} {121}},\ \bibinfo {pages} {147201} (\bibinfo {year}
  {2018})}\BibitemShut {NoStop}%
\bibitem [{\citenamefont {Bode}(2003)}]{Bode2003_stm}%
  \BibitemOpen
  \bibfield  {author} {\bibinfo {author} {\bibfnamefont {M.}~\bibnamefont
  {Bode}},\ }\bibfield  {title} {\enquote {\bibinfo {title} {{Spin-polarized
  scanning tunnelling microscopy}},}\ }\href {\doibase
  10.1088/0034-4885/66/4/203} {\bibfield  {journal} {\bibinfo  {journal}
  {Reports on Progress in Physics}\ }\textbf {\bibinfo {volume} {66}},\
  \bibinfo {pages} {523--582} (\bibinfo {year} {2003})}\BibitemShut {NoStop}%
\bibitem [{sup()}]{supplementary}%
  \BibitemOpen
  \href@noop {} {}\bibinfo {howpublished} {{see supplementary
  material}}\BibitemShut {NoStop}%
\bibitem [{\citenamefont {{Katsura, H. and Nagaosa, N. and Lee, P.
  A.}}(2010)}]{Katsura2010_magnonHall}%
  \BibitemOpen
  \bibfield  {author} {\bibinfo {author} {\bibnamefont {{Katsura, H. and
  Nagaosa, N. and Lee, P. A.}}},\ }\bibfield  {title} {\enquote {\bibinfo
  {title} {{Theory of the Thermal Hall Effect in Quantum Magnets}},}\ }\href
  {\doibase 10.1103/PhysRevLett.104.066403} {\bibfield  {journal} {\bibinfo
  {journal} {Phys. Rev. Lett.}\ }\textbf {\bibinfo {volume} {104}},\ \bibinfo
  {pages} {066403} (\bibinfo {year} {2010})}\BibitemShut {NoStop}%
\bibitem [{\citenamefont {Zhang}\ \emph {et~al.}(2013)\citenamefont {Zhang},
  \citenamefont {Ren}, \citenamefont {Wang},\ and\ \citenamefont
  {Li}}]{Zhang2013_magnons}%
  \BibitemOpen
  \bibfield  {author} {\bibinfo {author} {\bibfnamefont {L.}~\bibnamefont
  {Zhang}}, \bibinfo {author} {\bibfnamefont {J.}~\bibnamefont {Ren}}, \bibinfo
  {author} {\bibfnamefont {J.-S.}\ \bibnamefont {Wang}}, \ and\ \bibinfo
  {author} {\bibfnamefont {B.}~\bibnamefont {Li}},\ }\bibfield  {title}
  {\enquote {\bibinfo {title} {{Topological magnon insulator in insulating
  ferromagnet}},}\ }\href {\doibase 10.1103/PhysRevB.87.144101} {\bibfield
  {journal} {\bibinfo  {journal} {Phys. Rev. B}\ }\textbf {\bibinfo {volume}
  {87}},\ \bibinfo {pages} {144101} (\bibinfo {year} {2013})}\BibitemShut
  {NoStop}%
\bibitem [{\citenamefont {Malz}\ \emph {et~al.}(2019)\citenamefont {Malz},
  \citenamefont {Knolle},\ and\ \citenamefont
  {Nunnenkamp}}]{malz2019topological}%
  \BibitemOpen
  \bibfield  {author} {\bibinfo {author} {\bibfnamefont {D.}~\bibnamefont
  {Malz}}, \bibinfo {author} {\bibfnamefont {J.}~\bibnamefont {Knolle}}, \ and\
  \bibinfo {author} {\bibfnamefont {A.}~\bibnamefont {Nunnenkamp}},\ }\bibfield
   {title} {\enquote {\bibinfo {title} {{Topological magnon amplification}},}\
  }\href {\doibase 10.1038/s41467-019-11914-2} {\bibfield  {journal} {\bibinfo
  {journal} {Nature Communications}\ }\textbf {\bibinfo {volume} {10}}
  (\bibinfo {year} {2019}),\ 10.1038/s41467-019-11914-2}\BibitemShut {NoStop}%
\bibitem [{\citenamefont {Knolle}\ \emph {et~al.}(2014)\citenamefont {Knolle},
  \citenamefont {Kovrizhin}, \citenamefont {Chalker},\ and\ \citenamefont
  {Moessner}}]{Knolle2014_spinliquid}%
  \BibitemOpen
  \bibfield  {author} {\bibinfo {author} {\bibfnamefont {J.}~\bibnamefont
  {Knolle}}, \bibinfo {author} {\bibfnamefont {D.~L.}\ \bibnamefont
  {Kovrizhin}}, \bibinfo {author} {\bibfnamefont {J.~T.}\ \bibnamefont
  {Chalker}}, \ and\ \bibinfo {author} {\bibfnamefont {R.}~\bibnamefont
  {Moessner}},\ }\bibfield  {title} {\enquote {\bibinfo {title} {{Dynamics of a
  Two-Dimensional Quantum Spin Liquid: Signatures of Emergent Majorana Fermions
  and Fluxes}},}\ }\href {\doibase 10.1103/PhysRevLett.112.207203} {\bibfield
  {journal} {\bibinfo  {journal} {Phys. Rev. Lett.}\ }\textbf {\bibinfo
  {volume} {112}},\ \bibinfo {pages} {207203} (\bibinfo {year}
  {2014})}\BibitemShut {NoStop}%
\bibitem [{\citenamefont {{Knolle, J. and Kovrizhin, D. L. and Chalker, J. T.
  and Moessner, R.}}(2015)}]{knolle_2015_fract}%
  \BibitemOpen
  \bibfield  {author} {\bibinfo {author} {\bibnamefont {{Knolle, J. and
  Kovrizhin, D. L. and Chalker, J. T. and Moessner, R.}}},\ }\bibfield  {title}
  {\enquote {\bibinfo {title} {{Dynamics of fractionalization in quantum spin
  liquids}},}\ }\href {\doibase 10.1103/PhysRevB.92.115127} {\bibfield
  {journal} {\bibinfo  {journal} {Phys. Rev. B}\ }\textbf {\bibinfo {volume}
  {92}},\ \bibinfo {pages} {115127} (\bibinfo {year} {2015})}\BibitemShut
  {NoStop}%
\bibitem [{\citenamefont {{Knolle, J.}}(2016)}]{knolle_2016_dynamics}%
  \BibitemOpen
  \bibfield  {author} {\bibinfo {author} {\bibnamefont {{Knolle, J.}}},\
  }\href@noop {} {\emph {\bibinfo {title} {{Dynamics of a Quantum Spin
  Liquid}}}}\ (\bibinfo  {publisher} {Springer},\ \bibinfo {year}
  {2016})\BibitemShut {NoStop}%
\bibitem [{\citenamefont {Baskaran}\ \emph {et~al.}(2007)\citenamefont
  {Baskaran}, \citenamefont {Mandal},\ and\ \citenamefont
  {Shankar}}]{baskaran2007exact}%
  \BibitemOpen
  \bibfield  {author} {\bibinfo {author} {\bibfnamefont {G.}~\bibnamefont
  {Baskaran}}, \bibinfo {author} {\bibfnamefont {Saptarshi}\ \bibnamefont
  {Mandal}}, \ and\ \bibinfo {author} {\bibfnamefont {R.}~\bibnamefont
  {Shankar}},\ }\bibfield  {title} {\enquote {\bibinfo {title} {Exact results
  for spin dynamics and fractionalization in the kitaev model},}\ }\href
  {\doibase 10.1103/PhysRevLett.98.247201} {\bibfield  {journal} {\bibinfo
  {journal} {Phys. Rev. Lett.}\ }\textbf {\bibinfo {volume} {98}},\ \bibinfo
  {pages} {247201} (\bibinfo {year} {2007})}\BibitemShut {NoStop}%
\bibitem [{\citenamefont {Carrega}\ \emph {et~al.}(2020)\citenamefont
  {Carrega}, \citenamefont {Vera-Marun},\ and\ \citenamefont
  {Principi}}]{carrega2020tunneling}%
  \BibitemOpen
  \bibfield  {author} {\bibinfo {author} {\bibfnamefont {M.}~\bibnamefont
  {Carrega}}, \bibinfo {author} {\bibfnamefont {I.~J.}\ \bibnamefont
  {Vera-Marun}}, \ and\ \bibinfo {author} {\bibfnamefont {A.}~\bibnamefont
  {Principi}},\ }\href@noop {} {\enquote {\bibinfo {title} {{Tunneling
  spectroscopy as a probe of fractionalization in 2D magnetic
  heterostructures}},}\ } (\bibinfo {year} {2020}),\ \Eprint
  {http://arxiv.org/abs/2004.13036} {arXiv:2004.13036} \BibitemShut {NoStop}%
\bibitem [{\citenamefont {Casola}\ \emph {et~al.}(2018)\citenamefont {Casola},
  \citenamefont {van~der Sar},\ and\ \citenamefont {Yacoby}}]{Casola2018_nv}%
  \BibitemOpen
  \bibfield  {author} {\bibinfo {author} {\bibfnamefont {F.}~\bibnamefont
  {Casola}}, \bibinfo {author} {\bibfnamefont {T.}~\bibnamefont {van~der Sar}},
  \ and\ \bibinfo {author} {\bibfnamefont {A.}~\bibnamefont {Yacoby}},\
  }\bibfield  {title} {\enquote {\bibinfo {title} {{Probing condensed matter
  physics with magnetometry based on nitrogen-vacancy centres in diamond}},}\
  }\href {\doibase 10.1038/natrevmats.2017.88} {\bibfield  {journal} {\bibinfo
  {journal} {Nature Reviews Materials}\ }\textbf {\bibinfo {volume} {3}}
  (\bibinfo {year} {2018}),\ 10.1038/natrevmats.2017.88}\BibitemShut {NoStop}%
\bibitem [{\citenamefont {Pedrocchi}\ \emph {et~al.}(2011)\citenamefont
  {Pedrocchi}, \citenamefont {Chesi},\ and\ \citenamefont
  {Loss}}]{Pedrocchi2011_physical}%
  \BibitemOpen
  \bibfield  {author} {\bibinfo {author} {\bibfnamefont {F.~L.}\ \bibnamefont
  {Pedrocchi}}, \bibinfo {author} {\bibfnamefont {S.}~\bibnamefont {Chesi}}, \
  and\ \bibinfo {author} {\bibfnamefont {D.}~\bibnamefont {Loss}},\ }\bibfield
  {title} {\enquote {\bibinfo {title} {{Physical solutions of the Kitaev
  honeycomb model}},}\ }\href {\doibase 10.1103/PhysRevB.84.165414} {\bibfield
  {journal} {\bibinfo  {journal} {Phys. Rev. B}\ }\textbf {\bibinfo {volume}
  {84}},\ \bibinfo {pages} {165414} (\bibinfo {year} {2011})}\BibitemShut
  {NoStop}%
\bibitem [{\citenamefont {Zschocke}\ and\ \citenamefont
  {Vojta}(2015)}]{Zschocke2015_physical}%
  \BibitemOpen
  \bibfield  {author} {\bibinfo {author} {\bibfnamefont {F.}~\bibnamefont
  {Zschocke}}\ and\ \bibinfo {author} {\bibfnamefont {M.}~\bibnamefont
  {Vojta}},\ }\bibfield  {title} {\enquote {\bibinfo {title} {{Physical states
  and finite-size effects in Kitaev's honeycomb model: Bond disorder, spin
  excitations, and NMR line shape}},}\ }\href {\doibase
  10.1103/PhysRevB.92.014403} {\bibfield  {journal} {\bibinfo  {journal} {Phys.
  Rev. B}\ }\textbf {\bibinfo {volume} {92}},\ \bibinfo {pages} {014403}
  (\bibinfo {year} {2015})}\BibitemShut {NoStop}%
\end{thebibliography}%

\newpage \leavevmode \newpage
\appendix

\onecolumngrid

\section{\large{Supplementary Material}}
\begin{center}
\textbf{Local Probes for Charge-Neutral Edge States in Two Dimensional Quantum Magnets}\\ \vspace{10pt}
Johannes Feldmeier$^{1,2}$, Willian Natori$^{3}$, Michael Knap$^{1,2}$, Johannes Knolle$^{1,2,3}$ \\ \vspace{6pt}

$^1$\textit{\small{Department of Physics and Institute for Advanced Study, Technical University of Munich, 85748 Garching, Germany}} \\
$^2$\textit{\small{Munich Center for Quantum Science and Technology (MCQST), Schellingstr. 4, D-80799 M{\"u}nchen, Germany}} \\
$^3$\textit{\small{Blackett Laboratory, Imperial College London, London SW7 2AZ, United Kingdom}}
\end{center}
\maketitle

\section{1. Derivation of STM conductance}
Here, we provide some details on the derivation of \eq{eq:s1}. The following is essentially a mix of the derivations presented in Refs.~\cite{Rossier2009_stm,Balatsky2010_stm}. Let us describe the tripartite system laid out in the main text in terms of the eigenstates $\ket{\Psi} := \ket{n}_S\ket{\phi}_t\ket{\psi}_s$ of its three unperturbed constituents with respective energies $E_\Psi = E^S_n+E^t_\phi+E^s_\psi$. The experimentally relevant tunneling current $I$ between tip and substrate at inverse temperature $\beta$ can then be obtained most directly by applying Fermi's golden rule,
\begin{equation} \label{eq:A1.1}
\begin{split}
I = \frac{2e}{\hbar}\sum_{\bs{p},\bs{k},\sigma,\sigma^\prime}\sum_{\Psi,\tilde{\Psi}} e^{-\beta E_\psi}
\Bigl\{ \bigl| \bra{\tilde{\Psi}} \hat{T}^{\sigma\sigma^\prime}_{\bs{r}}\hat{a}^\dagger_{\bs{p},\sigma}\hat{b}^{}_{\bs{k},\sigma^\prime} \ket{\Psi} \bigr|^2 \; \delta(E_{\tilde{\Psi}}-E_\Psi -eV) - \bigl| \bra{\tilde{\Psi}} (\hat{T}^{\sigma\sigma^\prime}_{\bs{r}})^\dagger \hat{b}^\dagger_{\bs{k},\sigma^\prime} \hat{a}^{}_{\bs{p},\sigma} \ket{\Psi} \bigr|^2 \; \delta(E_{\tilde{\Psi}}-E_\Psi +eV) \Bigr\}.
\end{split}
\end{equation}
\eq{eq:A1.1} consists of two terms which we are going to treat seperately. The evaluation of the first matrix element can be decomposed into electron and spin sector via
\begin{equation} \label{eq:A1.2}
e^{-\beta E_\Psi}\bigl| \bra{\tilde{\Psi}} \hat{T}^{\sigma\sigma^\prime}_{\bs{r}}\hat{a}^\dagger_{\bs{p},\sigma}\hat{b}^{}_{\bs{k},\sigma^\prime} \ket{\Psi} \bigr|^2 = e^{-\beta E^S_n}\bigl|\braket{m| \hat{T}^{\sigma\sigma^\prime}_{\bs{r}}|n}\bigr|^2 \; e^{-\beta (E^t_\phi + E^s_\psi)}\bigl|\braket{\tilde{\phi},\tilde{\psi}|\hat{a}^\dagger_{\bs{p},\sigma}\hat{b}^{}_{\bs{k},\sigma^\prime}|\phi,\psi}\bigr|^2.
\end{equation}
Furthermore, due to the non-interacting nature of the metallic tip and substrate, the on-shell condition becomes $\delta(E_{\tilde{\Psi}}-E_\Psi -eV) = \delta(E^S_m-E^S_n+\varepsilon_{\bs{p}}-\varepsilon_{\bs{k}}-eV)$. As this does not explicitly depend on  $\tilde{\phi},\tilde{\psi},\phi,\psi$, we can carry out the corresponding summations in \eq{eq:A1.1}, i.e.
\begin{equation} \label{eq:A1.3}
\begin{split}
&\sum_{\phi,\psi}\sum_{\tilde{\phi},\tilde{\psi}} e^{-\beta (E^t_\phi + E^s_\psi)}\bigl|\braket{\tilde{\phi},\tilde{\psi}|\hat{a}^\dagger_{\bs{p},\sigma}\hat{b}^{}_{\bs{k},\sigma^\prime}|\phi,\psi}\bigr|^2 = \sum_{\phi,\psi}\sum_{\tilde{\phi},\tilde{\psi}} e^{-\beta (E^t_\phi + E^s_\psi)} \braket{\phi,\psi|\hat{b}^\dagger_{\bs{k},\sigma^\prime} \hat{a}^{}_{\bs{p},\sigma}|\tilde{\phi},\tilde{\psi}}\braket{\tilde{\phi},\tilde{\psi}|\hat{a}^\dagger_{\bs{p},\sigma}\hat{b}^{}_{\bs{k},\sigma^\prime}|\phi,\psi} = \\
&= \sum_{\phi,\psi} e^{-\beta (E^t_\phi + E^s_\psi)} \braket{\phi,\psi|\hat{b}^\dagger_{\bs{k},\sigma^\prime} \hat{b}^{}_{\bs{k},\sigma^\prime} \hat{a}^{}_{\bs{p},\sigma}\hat{a}^\dagger_{\bs{p},\sigma}|\phi,\psi} = \braket{\hat{b}^\dagger_{\bs{k},\sigma^\prime} \hat{b}^{}_{\bs{k},\sigma^\prime}}_\beta \braket{\hat{a}^{}_{\bs{p},\sigma}\hat{a}^\dagger_{\bs{p},\sigma}}_\beta = f(\varepsilon_{\bs{k}})(1-f(\varepsilon_{\bs{p}})),
\end{split}
\end{equation}
where $f(\varepsilon)$ is the Fermi distribution function at a given inverse temperature. We then proceed by converting the momentum summations $\sum_{\bs{p},\sigma}\rightarrow \sum_\sigma \int d\varepsilon \;n_\sigma(\varepsilon)$, $\sum_{\bs{k},\sigma}\rightarrow \sum_\sigma \int d\varepsilon \;N_\sigma(\varepsilon)$ into integrals over the densities of states $n_\sigma(\varepsilon),N_\sigma(\varepsilon)$ of tip and substrate electrons. We further assume that only electrons near the Fermi level contribute to tunneling, thus setting the densities of states $n_\sigma(\varepsilon)\rightarrow n_\sigma(\varepsilon_F),N_\sigma(\varepsilon)\rightarrow N_\sigma(\varepsilon_F)$ constant. Inserting this and \eq{eq:A1.3} into \eq{eq:A1.1} we obtain for the first term:
\begin{equation} \label{eq:A1.4}
\begin{split}
&\sum_{\bs{p},\bs{k},\sigma,\sigma^\prime}\sum_{\Psi,\tilde{\Psi}} e^{-\beta E_\psi}
\bigl| \bra{\tilde{\Psi}} \hat{T}^{\sigma\sigma^\prime}_{\bs{r}}\hat{a}^\dagger_{\bs{p},\sigma}\hat{b}^{}_{\bs{k},\sigma^\prime} \ket{\Psi} \bigr|^2 \; \delta(E_{\tilde{\Psi}}-E_\Psi -eV) = \\
&= \sum_{\sigma,\sigma^\prime} n_\sigma(\varepsilon_F) N_{\sigma^\prime}(\varepsilon_F) \sum_{n,m}e^{-\beta E^S_n} \bigl|\braket{m| \hat{T}^{\sigma\sigma^\prime}_{\bs{r}}|n}\bigr|^2 \int d\varepsilon \, d\varepsilon^\prime f(\varepsilon^\prime)(1-f(\varepsilon))\, \delta(E^S_m-E^S_n+\varepsilon-\varepsilon^\prime-eV) =\\ 
&=\sum_{\sigma,\sigma^\prime} n_{\sigma}(\varepsilon_F) N_{\sigma^\prime}(\varepsilon_F) \sum_{n,m}e^{-\beta E^S_n} \bigl|\braket{m| \hat{T}^{\sigma\sigma^\prime}_{\bs{r}}|n}\bigr|^2 \frac{eV-(E^S_m-E^S_n)}{1-e^{-\beta(eV-(E^S_m-E^S_n))}},
\end{split}
\end{equation}
where we carried out the integrals over $d\varepsilon, d\varepsilon^\prime$ in the last step.\\

We now evaluate the remaining summations over the spin sector. Firstly, we find for the tunneling matrix element, concentrating exclusively on the contributions $\sim t_1^2$ due to spin fluctuations,
\begin{equation} \label{eq:A1.5}
\bigl| \braket{m|\hat{T}^{\sigma\sigma^\prime}_{\bs{r}}|n} \bigr|^2 = \sum_{i,j}\sum_{\alpha,\beta} t_1(\bs{r}-\bs{r}_i) \, t_1(\bs{r}-\bs{r}_j)\; \sigma^\alpha_{\sigma^\prime \sigma} \sigma^\beta_{\sigma \sigma^\prime} \braket{n|\hat{S}^\alpha_i|m}\braket{m|\hat{S}^\beta_j|n}.
\end{equation}
We can then use the Lehmann representation of the Fourier transformed dynamical structure factor
\begin{equation} \label{eq:A1.6}
\mathcal{S}^{\alpha\beta}_{ij}(\omega) = \int dt e^{i\omega t} \braket{\hat{S}^\alpha_i(t)\hat{S}^\beta_j(0)} = \sum_{n,m} e^{-\beta E^S_n} \braket{n|\hat{S}^\alpha_i|m}\braket{m|\hat{S}^\beta_j|m}\, \delta(\omega - (E^S_m-E^S_n))
\end{equation}
to realize that for an arbitrary function $F(\omega)$, the following relation holds:
\begin{equation} \label{eq:A1.7}
\int d\omega\, \mathcal{S}^{\alpha\beta}_{ij}(\omega)\, F(\omega) = \sum_{n,m} e^{-\beta E^S_n} \braket{n|\hat{S}^\alpha_i|m}\braket{m|\hat{S}^\beta_j|n}\, F(E^S_m-E^S_n).
\end{equation}
Using this relation upon inserting the matrix element \eq{eq:A1.5} back into \eq{eq:A1.4}, we obtain for the first term of \eq{eq:A1.1},
\begin{equation} \label{eq:A1.8}
\begin{split}
&\sum_{\bs{p},\bs{k},\sigma,\sigma^\prime}\sum_{\Psi,\tilde{\Psi}} e^{-\beta E_\psi}
\bigl| \bra{\tilde{\Psi}} \hat{T}^{\sigma\sigma^\prime}_{\bs{r}}\hat{a}^\dagger_{\bs{p},\sigma}\hat{b}^{}_{\bs{k},\sigma^\prime} \ket{\Psi} \bigr|^2 \; \delta(E_{\tilde{\Psi}}-E_\Psi -eV) = \\
&= \sum_{i,j}\sum_{\alpha,\beta} t_1(\bs{r}-\bs{r}_i)t_1(\bs{r}-\bs{r}_j) \Bigl(\sum_{\sigma,\sigma^\prime} n_\sigma(\varepsilon_F)N_{\sigma^\prime}(\varepsilon_F) \sigma^{\alpha}_{\sigma^\prime \sigma}\sigma^\beta_{\sigma\sigma^\prime} \Bigr) \int d\omega \; \frac{eV-\omega}{1-e^{-\beta(eV-\omega)}}\; \mathcal{S}^{\alpha\beta}_{ij}(\omega)\\
&\rightarrow \sum_{i,j}\sum_{\alpha} t_1(\bs{r}-\bs{r}_i)t_1(\bs{r}-\bs{r}_j)\, c_{\alpha\beta} \int d\omega \; \frac{eV-\omega}{1-e^{-\beta(eV-\omega)}}\; \mathcal{S}^{\alpha\alpha}_{ij}(\omega),
\end{split}
\end{equation}
where in the last step we identified the weight function $c_{\alpha\beta}$ from the main text.\\

Repeating the same steps for the second term of the Fermi golden rule expression, we eventually arrive at the final expression for the current
\begin{equation} \label{eq:A1.9}
\begin{split}
I=\frac{2e}{\hbar}\sum_{i,j,\alpha}t_1(\bs{r}-\bs{r}_i)t_1(\bs{r}-\bs{r}_j)\,
 c_\alpha \int d\omega \, j_{V}(\omega)\, \mathcal{S}_{ij}^{\alpha\alpha}(\omega),
\end{split}
\end{equation}
\eq{eq:A1.9} contains the frequency weight function
\begin{equation} \label{eq:A.10}
j_{V}(\omega) = \frac{eV-\omega}{1-e^{-\beta(eV-\omega)}} + \frac{eV+\omega}{1-e^{\beta(eV+\omega)}},
\end{equation}
which reduces to $j_V(\omega)=(eV-\omega)\,\theta(eV-\omega)$ at zero temperature. Derivation of \eq{eq:A1.9} with respect to $V$ yields \eq{eq:s1} of the main text.

\section{2. Kitaev Honeycomb Model}
We provide further information and details on the computation of the dynamical structure factor in the extended Kitaev model on open boundaries. Particular attention is devoted to the subtleties arising from ground state degeneracies in the OBC limit.
\subsubsection{Physical Hilbert space}
The decomposition of a spin-$1/2$ into four Majoranas introduced by Kitaev enlarges the Hilbert space. The projection back onto the physical Hilbert space is obtained by requiring that $\hat{D}_i=-i \hat{\sigma}^x_i \hat{\sigma}^y_i \hat{\sigma}^z_i = \hat{b}^x_i \hat{b}^y_i \hat{b}^z_i \hat{c}_i=1$ for all sites. This condition can be enforced in terms of the bond and matter fermions via the projection operator
\begin{equation} \label{eq:2.8}
\hat{P} = \prod_i \frac{1+\hat{D}_i}{2} \sim \frac{1}{2}\left[1 + (-1)^{N_f+N_\chi}\right],
\end{equation}
where $N_f/N_\chi$ are the total number of matter/bond fermions. \eq{eq:2.8} demonstrates that only states with even total fermion number parity lie within the physical spin Hilbert space. As was shown in Refs.~\cite{Pedrocchi2011_physical, Zschocke2015_physical}, particular care needs to be taken within the gapless phase of the pure Kitaev model when projecting back to the physical Hilbert space.

\subsection{Open boundaries}
As outlined in the main text, open boundary conditions can be obtianed by introducing a line of `weak bonds' as shown in \fig{fig:A1}, where all terms in the Hamiltonian \eq{eq:s2} involving such bonds are multiplied by a factor $J_b<1$. The case of open boundaries is then retrieved for $J_b=0$, which effectively cuts the system in half. For the practical evaluation of structure factors, we choose the value of the weak bonds very small, $J_b\ll 1$, but finite. This allows us to directly use the numerical method derived for periodic boundaries ~\cite{knolle_2015_fract,knolle_2016_dynamics}. In practice, we work on a cylindrical geometry, and neglect a non-local ground state degeneracy due to invariant Wilson loops winding around the cylinder, which does not affect our local probe results.

However, we emphasize that one has to be careful when taking the limit $J_b\rightarrow 0$. We discuss in the following how this limit impacts both the ground state structure as well as the dynamical spin correlations.

\subsubsection{Ground state degeneracy: Gauge sector}
\begin{figure}[t]
\begin{center}
\includegraphics[trim={0cm 0cm 0cm 0cm},clip,width=.5\linewidth]{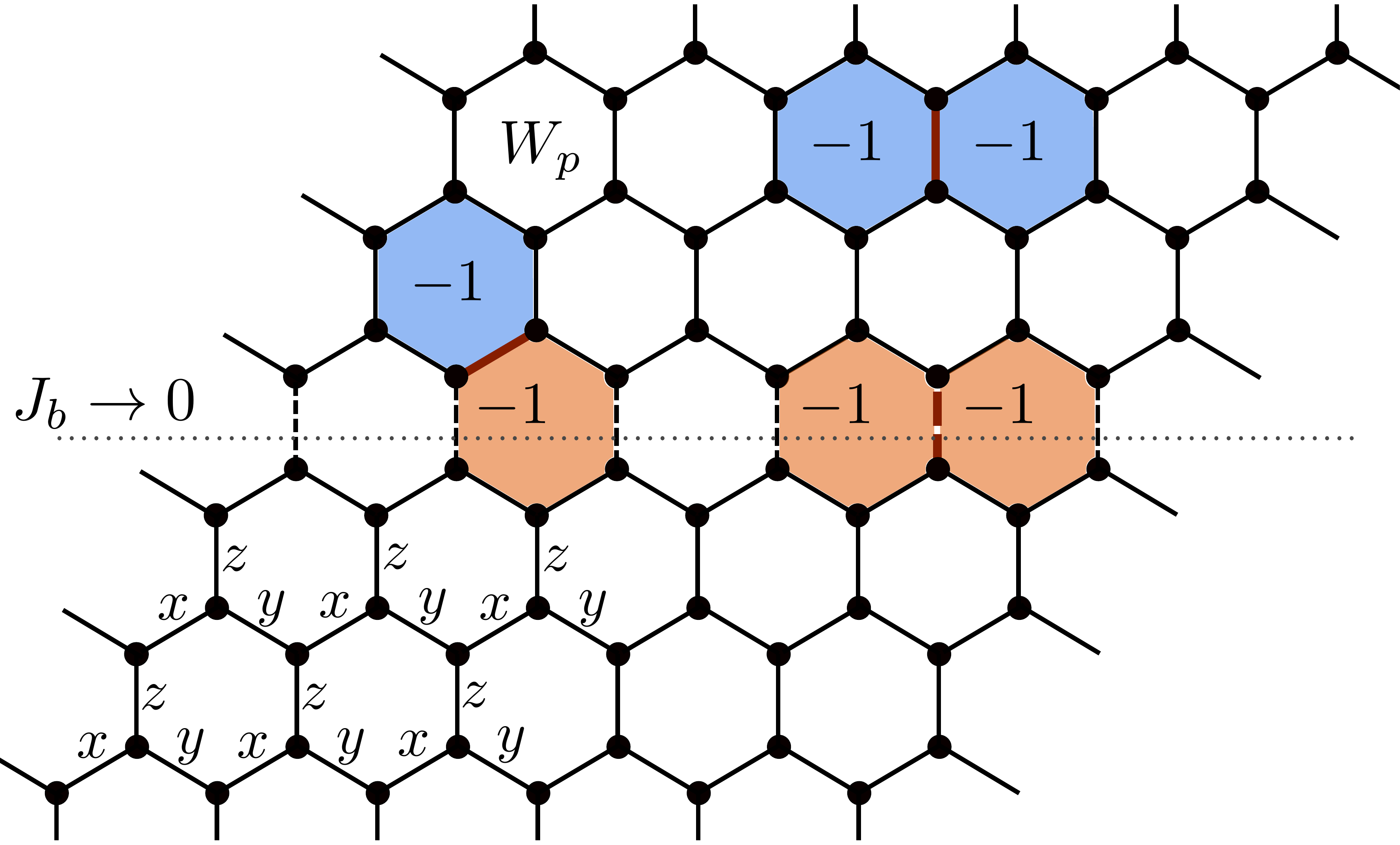}
\caption{\textbf{Kitaev model.} The three types of bonds are labelled according to the anisotropic exchange interaction of the Hamiltonian \eq{eq:s2} of the main text. The interactions along a line of $z$-bonds through the system are weakened by a factor $J_b<1$, yielding open boundary conditions for $J_b=0$. Inserting bond fermions (red bonds) flips the flux $W_p$ of the two adjacent plaquettes. In the bulk, these flux excitations are gapped (blue plaquettes), while boundary plaquettes cost no energy and lead to degeneracies in the spectrum (orange plaquettes).} 
 \label{fig:A1}
\end{center}
\end{figure}

As discussed above, the ground state of the translationally invariant system $J_b=1$ is unique and lies in the sector of zero flux. This property remains true for any non-zero $J_b>0$, for which the minimal flux gap is of order $\sim (J_b \cdot J)$, a property we have verified numerically on finite size systems. However, for $J_b=0$ exactly, plaquette fluxes adjacent to the weak bonds can be inserted at the newly formed system boundary without energy cost. Formally, if we let $\braket{ij}_b$ denote one of the weak bonds as shown in \fig{fig:A1}, this can be expressed via $[\hat{\chi}_{\braket{ij}_b},\hat{H}]=0$. We notice however that in order to obtain a valid transformation within the physical Hilbert space that respects the parity selection rule of \eq{eq:2.8}, we need to create/annihilate an \textit{even} number of boundary gauge fermions, starting from the original flux-free ground state. The set of transformations that relate different ground states is thus given by
\begin{equation} \label{eq:2.15}
u_{\braket{ij}_b} \rightarrow -u_{\braket{ij}_b}, \quad  u_{\braket{kl}_b} \rightarrow -u_{\braket{kl}_b},
\end{equation}
for an arbitrary pair of boundary bonds $\braket{ij}_b$,$\braket{kl}_b$. From this we can infer the total ground state degeneracy $D_{f}$ due to boundary fluxes for a system of linear length $L$ along the open boundary to be 
\begin{equation} \label{eq:2.16}
D = \binom{L}{0} + \binom{L}{2} + \binom{L}{4} + ... = 2^{L-1}.
\end{equation}
We have observed this degeneracy due to boundary fluxes using exact diagonalization methods for the original spin Hamiltonian \eq{eq:s2} on small system sizes. We notice further that this degeneracy applies to \textit{all} eigenenergies throughout the entire many body spectrum.

We can now write down the form of a general state within this degenerate manifold. The gauge sector will then be flux-free in the bulk and consist of a general superposition of fluxes on the boundary, leading to \eq{eq:s4} of the main text,
\begin{equation} \label{eq:2.17}
\ket{0} = \ket{M_0} \otimes \ket{F_0}_{bulk} \otimes \ket{F}_{b},
\end{equation}
with $\ket{F}_b$ a linear superposition of different boundary flux configurations.

\subsubsection{Ground state degeneracy: Matter sector}
As demonstrated in Kitaev's original work~\cite{kitaev_2006}, the energy bands of the matter fermions carry non-trivial Chern number for non-zero K, which implies the existence of chiral edge states within the bulk gap and a zero energy edge mode on open boundary conditions. An example was given directly in the Appendix of~\cite{kitaev_2006}. We notice that on finite systems, the mode with zero energy might not be directly visible, as the exact momentum hosting it might not be part of the reciprocal lattice. However, in the thermodynamic limit we are guaranteed the existence of $\ket{\tilde{M}_0} = \hat{a}^\dagger_0 \ket{M_0}$ with $E(\tilde{M}_0)=E(M_0)$.

Since $\ket{\tilde{M}_0}$ contains a matter fermion, we are now required to add an odd number of gauge fermions to obtain a physical state. In order to remain in a ground state, we add an odd number of \textit{boundary} gauge fermions, for which there are in turn again
\begin{equation} \label{eq:2.18}
\tilde{D} = \binom{L}{1} + \binom{L}{3} + \binom{L}{5} + ... = 2^{L-1}
\end{equation}
different possibilities. A general ground state within this matter sector is then given as
\begin{equation} \label{eq:2.19}
\ket{\tilde{0}} = \ket{\tilde{M}_0}\otimes \ket{F_0}_{bulk} \otimes \ket{\tilde{F}},
\end{equation}
with $\ket{\tilde{F}}$ a superposition of $\tilde{D}$ boundary flux sectors.\\

Taken together both matter and gauge sources of degeneracy, we obtain the total ground state degeneracy to be $2^L$-fold.

\subsubsection{Open boundaries: Structure factor}
After this detailed discussion of the open boundary limit $J_b=0$ in terms of ground state degeneracies, we wish to know how these results merge with our numerical approach of setting $J_b\ll 1$ but finite. In particular we would like to discuss how the dynamical structure factor differs between the unique ground state for $J_b>0$ and a general ground state for $J_b=0$ which is a superposition of $2^L$ different states from a degenerate manifold. Remarkably, while in general differences between the two cases do occur, the dominant on-site contribution relevant for the STM response will turn out to be independent of the chosen ground state, such that the limit $J_b\rightarrow 0$ is indeed continuous for the on-site spin correlations.

Let us take the system to be in one of the ground states $\ket{0}$ from \eq{eq:s4} and consider two sites $i,j\in A$ which are \textit{both} located on the boundary. We assume further, that the weak bonds that were removed in order to obtain open boundaries are $z$-bonds. We then compute the corresponding structure factor, using \eq{eq:s5} and the fact that $[\hat{\chi}^{}_{\braket{il}_b},\hat{H}]=0$ for boundary bonds, 
\begin{equation} \label{eq:2.20}
\begin{split}
\mathcal{S}^{zz}_{ij}& = \braket{M_0|e^{it\hat{H}}\hat{c}_ie^{-it\hat{H}}\hat{c}_j|M_0} \times \\
&\times \prescript{}{b}{\Braket{F|(\hat{\chi}^{}_{\braket{il}_b}+\hat{\chi}^\dagger_{\braket{il}_b})(\hat{\chi}^{}_{\braket{jk}_b}+\hat{\chi}^\dagger_{\braket{jk}_b})|F}_b}.
\end{split}
\end{equation}
Here, we have used that the bulk gauge sector remains unchanged, $\prescript{}{bulk}{\braket{F_0|F_0}_{bulk}}=1$. Because the boundary gauge sector $\ket{F}_b$ is now a general superposition, the expression \eq{eq:2.20} does \textit{not} reduce to an on-site contribution $\sim \delta_{ij}$ like in the periodic case~\cite{baskaran2007exact,Knolle2014_spinliquid}.

\begin{figure}[t]
\begin{center}
\includegraphics[trim={0cm 0cm 0cm 0cm},clip,width=.5\linewidth]{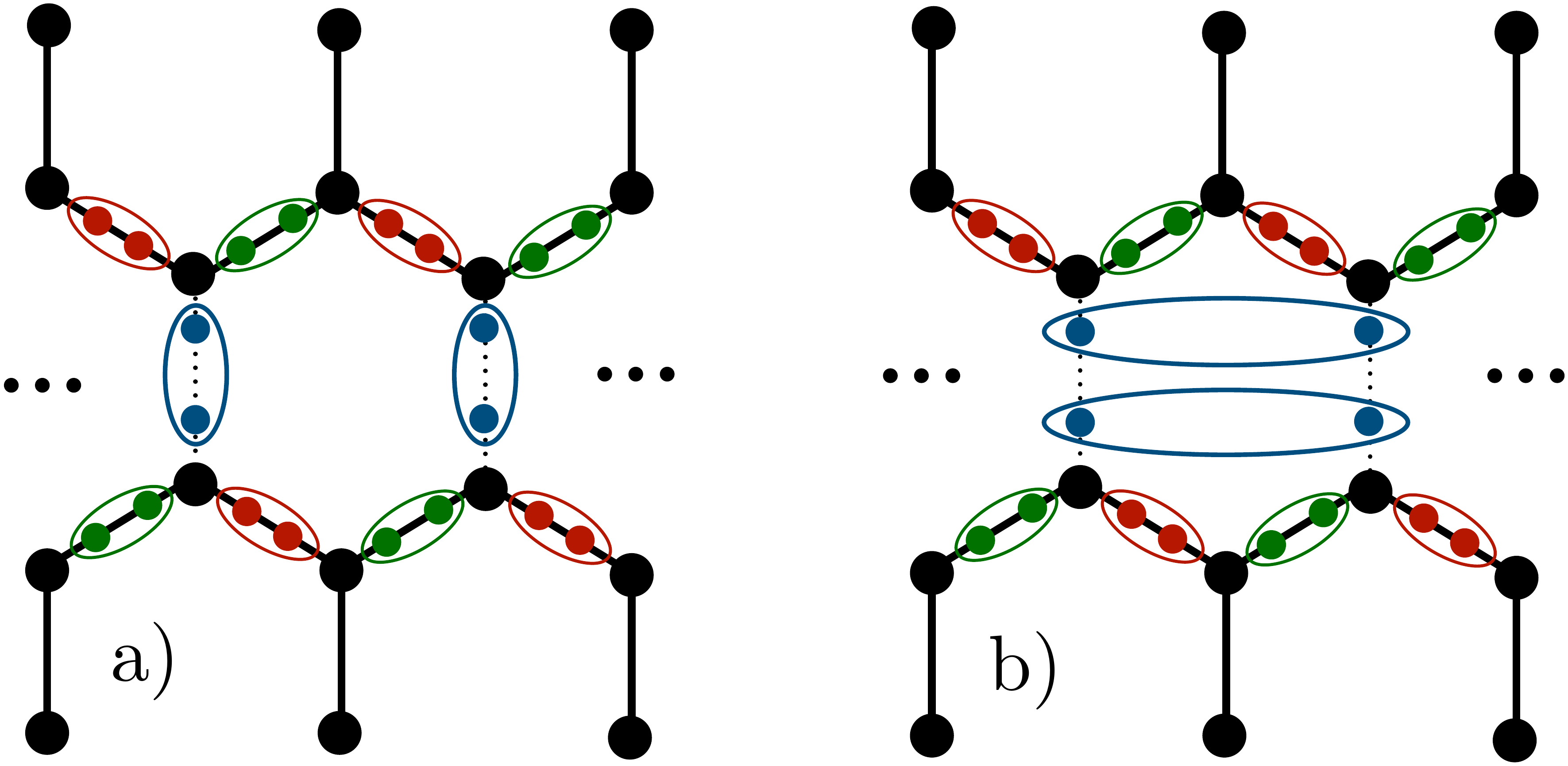}
\caption{\textbf{Majorana pairings:} The Majorana fermions $\hat{b}^\alpha_i$ (colored according to $\alpha$) are paired up to form the the gauge fermions $\hat{\chi}^{}_{ij}$ living on bonds $(i,j)$, whose occupation numbers commute with the Hamiltonian. If we introduce open boundaries by setting the exchange $J_b=0$ on the line of vertical bonds shown here (dotted bonds), there arises an \textit{ambiguity} in how to pair up the resulting `dangling' Majoranas (shown in blue). \textbf{a)} The original pairing along the former bond is still valid, and produces the usual ultra-local expression for the spin structure factor. \textbf{b)} The Marojanas can now also be paired up in longer-range bonds $(i,j)$, and the resulting fermion occupation numbers still commute with the Hamiltonian, allowing for longer-range contributions to the structure factor. The different pairings are related by a basis change within the degenerate ground state manifold.} 
 \label{fig:A2}
\end{center}
\end{figure}

An alternative way to see that there are indeed non-vanishing longer-range contributions beyond nearest neighbors to the structure factor for $J_b=0$ comes from `rewiring' the $\hat{b}^z_i$ - Majoranas on the boundary. As illustrated in \fig{fig:A2}, we can pair up the $\hat{b}^z_i$ in an arbitrary way to form new gauge fermions $\hat{\chi}_{(ij)_b}$, where $(ij)_b$ need not be lattice nearest neighbors. These new bond fermions still commute with the Hamiltonian and provide equally valid labellings of the model's gauge sector. Within this pairing, the new `nearest neighbors' can clearly provide non-vanishing spin correlations in full analogy to the previous nearest neighbor contributions derived in Ref~\cite{Knolle2014_spinliquid}. Thus, the rewiring of boundary Majoranas is equivalent to a basis change in the Fock space spanned by the occupation numbers $\hat{\chi}^\dagger_{\braket{ij}_b} \hat{\chi}^{}_{\braket{ij}_b}$.\\

While the spin correlations for off-diagonal site pairs $i\neq j$ are thus clearly dependent on the chosen ground state out of the degenerate manifold, we see that for on-site terms $i=j$ the flux part in \eq{eq:2.20} simplifies due to $(\hat{\chi}^{}_{\braket{il}_b}+\hat{\chi}^\dagger_{\braket{il}_b})(\hat{\chi}^{}_{\braket{ik}_b}+\hat{\chi}^\dagger_{\braket{ik}_b}) = \mathbb{1}$. We can thus conclude that the on-site structure factor is independent of the chosen state and
\begin{equation} \label{eq:2.21}
\begin{split}
\lim_{J_b\rightarrow 0} \left[\mathcal{S}^{\alpha\alpha}_{ii}(t)\big|_{J_b}\right] = \mathcal{S}^{\alpha\alpha}_{ii}(t)\big|_{J_b=0}.
\end{split}
\end{equation}
The limit $J_b \rightarrow 0$ is therefore indeed continuous for this contribution and couples directly to the on-site Majorana correlation function, providing an in principle even simpler expression than the quench problem that needs to be solved for bulk correlations. Furthermore, we do not expect \eq{eq:2.21} to change when including the degeneracy due to the zero energy matter mode $\ket{\tilde{M}_0}$: As the corresponding isolated mode is delocalized along the boundary, its effect on the local structure factor is expected to decrease as $\sim 1/L$ in system size. Furthermore, effects of finite temperature will smoothen out the response for $\omega \rightarrow 0$ in any case.

We have verified \eq{eq:2.21} independently on small finite size systems that can be treated with exact diagonalization or matrix product state techniques. The relation is convenient, as it allows us to draw direct conclusions about expected experimental signatures in open boundary conditions, while being able to formally work with the technical benefits of a periodic system.

\subsection{3. STM response: geometrical properties}
\begin{figure}[t]
\begin{center}
\includegraphics[trim={0cm 0cm 0cm 0cm},clip,width=.4\linewidth]{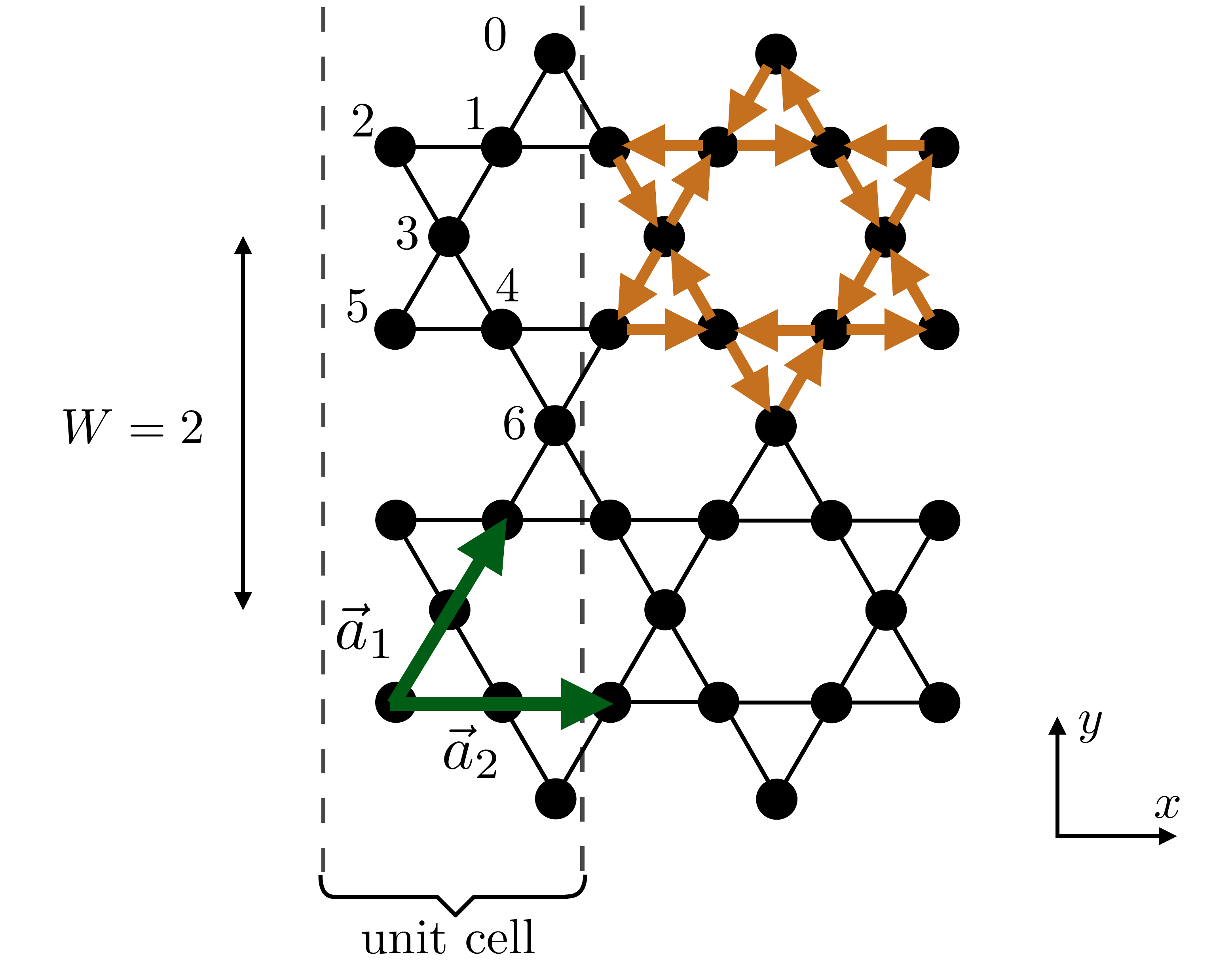}
\caption{\textbf{Geometry of the TMI setup.} (compare \figc{fig:s2}{a}) We consider open boundaries in $y$-direction, implying translational invariance only along the $\bs{a}_2$ lattice vector. The number of sites within a unit cell is $6W+1$; the position of a site $n$ is specified by $(x_n,l_n)$, with $x_n$ labelling the unit cell and $l_n\in \{0,...,6W\}$ labelling the site within a unit cell as depicted here.} 
 \label{fig:A3}
\end{center}
\end{figure}

We provide some more intuition on the dependence of the conductance on the geometry of the setup. In particular, for the example of the TMI in the main text, we considered a larger value of $\lambda \sim 1$ as the effective range of the exchange interactions entering $t_1(\bs{r}-\bs{r}_i) \sim e^{-|\bs{r}-\bs{r}_i|/\lambda}$.
Since the TMI system is block-diagonal with respect to the momentum $k_x$, we can work directly in an infinitely extended system in the $x$-direction using the Fourier transform $S^\alpha_n(t)=\frac{1}{\sqrt{L_x}}\sum_{k_x}e^{ik_xx_n}S^\alpha_{l_n}(k_x,t)$, where $x_n$ is the $x$-position of the kagome-site $n$, and $l_n\, \in \, \{0,...,6W\}$ determines the $y$-position within the unit cell as depicted in \fig{fig:A3}. We can then express the dynamical structure factor $\mathcal{S}^{\alpha\alpha}_{nm}$ in terms of its 1D Fourier transform according to
$\mathcal{S}^{\alpha\alpha}_{nm}(\omega)=\sum_{k_x} e^{ik_x(x_n-x_m)}\mathcal{S}^{\alpha\alpha}_{l_n l_m}(k_x,\omega)$.
Inserting into the expression \eq{eq:s1} for the conductance and using that $\sum_n \rightarrow \sum_{x_n,l_n}$ gives the simplified result
\begin{equation} \label{eq:3.6}
\begin{split}
\frac{\partial I}{\partial V} = \frac{2e^2}{\hbar} n(\varepsilon_F)N(\varepsilon_F) \times \sum_{l_n,l_m,k_x,\alpha} g^{}_{l_n\,l_m}(k_x,\bs{r}) \, c_\alpha \, \int_0^{eV} d\omega \, \mathcal{S}^{\alpha\alpha}_{l_nl_m}(k_x,\omega),
\end{split}
\end{equation}
with
\begin{equation} \label{eq:3.7}
\begin{split}
g^{}_{l_nl_m}(k_x,\bs{r}) = \Bigl( \sum_{x_n} e^{ik_x x_n}t_1(\bs{r}-\bs{r}_n)\Bigr) \times \Bigl( \sum_{x_n} e^{-ik_x x_m}t_1(\bs{r}-\bs{r}_m) \Bigr).
\end{split}
\end{equation}

It is instructive to approximate \eq{eq:3.7} by turning the sum into an integral and insert the form of $t_1(\bs{r}-\bs{r}_n)$ to obtain
\begin{equation} \label{eq:3.8}
\begin{split}
&\sum_{x_n} e^{ik_x x_n}t_1(\bs{r}-\bs{r}_n) \approx \int dx_n\, e^{ik_x x_n} t_1(\bs{r}-\bs{r}_n) =\Gamma_1\,e^{-d/d_0} \int dx_n\, e^{ik_x x_n} e^{-|\bs{r}-\bs{r}_n|/\lambda} = \\
&= 2\Gamma_1\,e^{-d/d_0} e^{ik_x x} \frac{|y-y_n|}{\sqrt{1+\lambda^2k_x^2}} \times K_1\left( \frac{|y-y_n|}{\lambda}\sqrt{1+\lambda^2k_x^2} \right),
\end{split}
\end{equation}
where $K_1(\cdot)$ is a modified Bessel function of the second kind and all lengths are measured in units of the lattice spacing. We notice further, that $y_n=y_n(l_n)$ is uniquely specified by the index $l_n\, \in \, \{0,...,6W\}$. With \eq{eq:3.8} at hand, the function $g^{}_{l_nl_m}(k_x,\bs{r})$ is determined and can be inserted back into \eq{eq:3.6}. $K_1(x)$ drops off exponentially for large arguments and diverges as $K_1(x)\sim 1/x$ for $x\rightarrow 0$, as would be relevant for e.g. the case $y=y_n$. We therefore see that the response acquired through the device function $g_{l_nl_m}(k_x,\bs{r})$ will only pick up sizeable contributions from momenta $k_x \lesssim 1/\lambda$. Importantly, the edge state in between the first and second energy band as displayed in \figc{fig:s2}{b} is located directly at $k_x=0$ and should therefore be able to contribute to the response as measured by the local conductance. This feature appears to arise for boundaries shaped differently than \fig{fig:A3} as well, see e.g. Ref~\cite{Zhang2013_magnons} for a $k_x=0$ edge state well separated in energy from the bulk.

\end{document}